 \documentclass[manuscript]{aastex63}

\usepackage[T1]{fontenc}
\usepackage{ae,aecompl}

\usepackage[version=4]{mhchem}
\usepackage{empheq}

\newenvironment{reactempheq}[2][]{%
  \renewcommand{\theequation}{R\arabic{equation}}
  \setkeys{EmphEqEnv}{#2}%
  \setkeys{EmphEqOpt}{#1}%
  \EmphEqMainEnv%
}{%
  \endEmphEqMainEnv%
}

\received{XXX X, 20XX}
\revised{XXX X, 20XX}
\accepted{April 17th, 2020}
\submitjournal{The Astrophysical Journal}

\shorttitle{Ions in the Thermosphere of Exoplanets}
\shortauthors{Bourgalais et al.}
\graphicspath{{./}{figures/}}

\begin{document}

\title{Ions in the Thermosphere of Exoplanets: {Observable} Constraints Revealed by Innovative Laboratory Experiments}

\correspondingauthor{J\'er\'emy Bourgalais}
\email{jeremy.bourgalais@latmos.ipsl.fr}

\author[0000-0003-4710-8943]{J\'er\'emy Bourgalais}
\affiliation{LATMOS/IPSL, UVSQ, Universit\'e Paris-Saclay, UPMC Univ. Paris 06, CNRS, Guyancourt, France.}

\author{Nathalie Carrasco}
\affiliation{LATMOS/IPSL, UVSQ, Universit\'e Paris-Saclay, UPMC Univ. Paris 06, CNRS, Guyancourt, France.}


\author{Quentin Changeat}
\affiliation{Department of Physics \& Astronomy, University College London, Gower Street, WC1E 6BT London, United Kingdom.}

\author{Olivia Venot}
\affiliation{Laboratoire Interuniversitaire des Syst\`emes Atmosph\'eriques (LISA), UMR CNRS 7583, Universit\'e Paris-Est-Cr\'eteil, Universit\'e de Paris, Institut Pierre Simon Laplace, Cr\'eteil, France.}

\author{Lora Jovanovi\'c}
\affiliation{LATMOS/IPSL, UVSQ, Universit\'e Paris-Saclay, UPMC Univ. Paris 06, CNRS, Guyancourt, France.}


\author{Pascal Pernot}
\affiliation{Laboratoire de Chimie Physique, CNRS, Univ. Paris-Sud, Universit\'e Paris-Saclay, 91405, Orsay, France.}

\author{Jonathan Tennyson}
\affiliation{Department of Physics \& Astronomy, University College London, Gower Street, WC1E 6BT London, United Kingdom.}

\author{Katy L. Chubb}
\affiliation{Department of Physics \& Astronomy, University College London, Gower Street, WC1E 6BT London, United Kingdom.}
\affiliation{SRON Netherlands Institute for Space Research, Sorbonnelaan 2, 3584 CA, Utrecht, Netherlands.}

\author{Sergey N. Yurchenko}
\affiliation{Department of Physics \& Astronomy, University College London, Gower Street, WC1E 6BT London, United Kingdom.}

\author{Giovanna Tinetti}
\affiliation{Department of Physics \& Astronomy, University College London, Gower Street, WC1E 6BT London, United Kingdom.}



\begin{abstract}
With the upcoming launch of space telescopes dedicated to the study of exoplanets, the \textit{Atmospheric Remote-Sensing Infrared Exoplanet Large-survey} (ARIEL) and the \textit{James Webb Space Telescope} (JWST), a new era is opening in the exoplanetary atmospheric explorations. However, especially in relatively cold planets around later-type stars, photochemical hazes and clouds may mask the composition of the lower part of the atmosphere, making it difficult to detect any chemical species in the troposphere or to understand whether there is a surface or not. This issue is particularly exacerbated if the goal is to study the habitability of said exoplanets and to search for biosignatures.\par 
This work combines innovative laboratory experiments, chemical modeling and simulated observations at ARIEL and JWST resolutions. We focus on the signatures of molecular ions that can be found in upper atmospheres above cloud decks. Our results suggest that H$_3^+$ along with H$_3$O$^+$ could be detected in the observational spectra of sub-Neptunes based on realistic mixing ratio assumption. This new parametric set may help to distinguish super-Earths with a thin atmosphere from H$_2$-dominated sub-Neptunes, to address the critical question whether a low-gravity planet around a low-mass active star is able to retain its volatile components. These ions may also constitute potential tracers to certain molecules of interest such as H$_2$O or O$_2$ to probe the habitability of exoplanets. Their detection will be an enthralling challenge for the future JWST and ARIEL telescopes.
\end{abstract}



\section{Introduction} \label{sec:intro}

Thanks to the {Kepler mission \citep{2016RPPh...79c6901B}}, many exoplanets of a new kind, bigger than the Earth but smaller than Neptune (1R$_{\oplus}$ $<$ R$_p$ $<$ 4R$_{\oplus}$) have been observed and constitute the majority of the worlds detected outside the solar system \citep{coughlin2016planetary}. Based on observations and theoretical considerations, many of these low-gravity planets seem to have thick atmospheres which retain light species (H$_2$, He) \citep{jontof2019compositional}. However, these small planets show a great diversity of densities, suggesting a diverse range of bulk and atmospheric compositions including rocky planets with thin atmospheres mainly made of heavy molecules such as N$_2$, CO, or H$_2$O \citep{rimmer2019hydrogen,hu2014photochemistry,moses2013compositional,elkins2008ranges,schaefer2012vaporization}. \par
These atmospheric constituents have a significant impact on the evolution of the planet, including its potential habitability. Many low-gravity planets are detected mainly in orbit around low-mass M-type stars which are the most abundant stars in our galaxy \citep{laughlin2004core,chabrier2003galactic}. These stars emit strongly in the UltraViolet (UV) and the atmospheric properties as well as the dynamics and fate of their associated exoplanets are thus strongly impacted by stellar radiation \citep{shkolnik2014hazmat}. Based on our current knowledge of the solar system's bodies, {such as Titan} \citep{horst2017titan}, high-energy radiation, in particular in the Extreme-UV (EUV) range (10 - 120 nm), has a strong effect on the chemical composition of the atmosphere \citep{chadney2015xuv}. Similar effects are therefore also expected in the atmospheres of exoplanets especially those orbiting M dwarfs which remain active in the EUV much longer than Sun-like G-stars (luminosity factors of 10 - 100 times higher than the solar flux \citep{ribas2016habitability}) \citep{lammer2011uv}. EUV radiation significantly alters the chemical composition of the upper atmosphere of exoplanets during the early phase of their evolution through interaction with the most abundant molecular species, but its long-term effect on the habitability of planets is complex and remains an open question \citep{luger2015extreme,wordsworth2014abiotic,bolmont2016water,owen2016habitability}. The thermosphere is a complex chemical environment where the main atmospheric components are dissociated and ionized, leading to efficient ion-molecule reactions involving their primary volatile elements \citep{moses2011disequilibrium}. Many atmospheric chemical models, adapted mostly to hot exoplanets, are available in the literature, some of which considering the effects of metals and simple ions, like H$_3^+$, in non-equilibrium models \citep{munoz2007physical,helling2019lightning,drummond2016effects,yelle2004aeronomy,lavvas2014electron,koskinen2013escape,rimmer2016chemical}. However, there is no existing study focused on the characterization of the main positive ions that may form within the thermosphere of warm exoplanets and their expected important effects on the chemistry are yet to be quantified. In particular, any effect on the distribution of the IR-radiating species which drive thermal and non-thermal atmospheric escape processes will influence the retention of planetary atmospheres. Determining whether a low-gravity planet around a small active star is able to retain its volatile components is a critical question to understand its nature and evolution. For rocky temperate planets, in particular, this question is pivotal to understand their potential habitability \citep{luger2015extreme,bourrier2017reconnaissance} since UV irradiation may also stimulate the synthesis of prebiotic molecules \citep{rimmer2018origin}.\par
We are at the dawn of a new era in space exploration with the upcoming launch of new telescopes like the \textit{Atmospheric Remote-Sensing Infrared Exoplanet Large-survey} (ARIEL) \citep{tinetti2018chemical} and the \textit{James Webb Space Telescope} (JWST) \citep{gardner2006james}. They will provide a unique insight into the thermal structure and chemical composition of the atmosphere of exoplanets, due to their spectral range and array of instruments that will be used for transmission, emission and phase-curve spectroscopy. To make the most of these future observations, it is important to learn as much as possible about their atmospheric chemistry. Among the different approaches, experimental laboratory simulation is a powerful tool for understanding the chemical evolution of an exoplanet atmosphere and also for guiding observations and the development of atmospheric models. However, so far in the literature, previous laboratory experiments have {focused on solid particles (production rates, size, chemical composition) that may constitute photochemical hazes within the cold and warm atmospheres of exoplanets \citep{horst2018haze,he2018laboratory,he2018photochemical,wolters2019orbitrap,moran2020chemistry,vuitton2019laboratory}}. \\
In order to complete this crucial phase of experimental simulation, it is also necessary to study the chemical evolution of the gas-phase in the primary upper atmosphere of exoplanets following UV irradiation of host stars and to identify the abundant species that may form. {While \citet{he2018gas} studied the neutral species involved in the gas-phase chemistry of simulated exoplanet atmospheres, the implication of the positive ions has never been studied.} Photoproducts drive the formation of hazes lower in the atmosphere but are also the starting point for the formation of prebiotic molecules. To be able to detect biosignatures that will be critical to the search for life, this work paves the first milestones with innovative photochemical-driven molecular growth laboratory experiments and shows the direct relationship with space observations. \par
{The approach in this article aims at providing information on chemical species, in particular, charged species that are expected in the thermosphere of rocky and gas-rich exoplanets and could detectable with future ARIEL and JWST observations. To achieve this goal and to show how the combination of laboratory simulation and numerical modeling constrains the inputs in the modeling of the observations, this work has been divided into three steps:\\
-In a first step, we have experimentally simulated the effect of UV radiation on simplified gas phase samples (H$_2$-CO-N$_2$-H$_2$O) whose variation of the hydrogen mixing ratio allows us to mimic cases of super-Earths and mini-Neptunes. This first experimental study allowed the measurements of both neutral and cations and to identify the main positive ions formed in H$_2$-poor and H$_2$-rich environments.\\
-In the second step, we analyzed the main pathways of formation of these ions using a photochemical model. The characterization of the photoproducts and the identification of key reactions allowed us to confirm that these experimental results were transposable to planetary environments and that these species could also be formed in significant abundance in the thermosphere of exoplanets. \\
-In the third step, we carried out simulations of observations using the resolution of the future JWST and ARIEL space telescopes to determine whether it was possible to detect in a transmission spectrum of an exoplanet two of the major ions observed in laboratory experiments with realistic mixing ratios. Due to the lack of chemical information on the atmospheres of exoplanets, we chose a planet whose composition most closely resembles one of the gaseous mixtures used in the laboratory (mini-Neptune GJ1214b) where CO replaced by methane is the only difference, but whose impact on the formation of ions is nil.\\}

\section{Methods} \label{sec:methods}

\subsection{Photochemical Reactor Coupled to an EUV Photon Source}

Simplified super-Earth and mini-Neptune atmospheres are reproduced in the laboratory using a photochemical reactor that has already been used in the past for photochemical experiments simulating complex environments similar to the upper layers of planetary atmospheres \citep{carrasco2013vuv,peng2014modeling,bourgalais2019low}. The cell is filled with a gas mixture representative of the environments to be reproduced. The species of the gas mixture are then irradiated by an EUV photon source windowless coupled to the photochemical reactor. The energetic particles are produced by a microwave plasma discharge of a rare gas at low pressure providing radiation at different wavelengths depending on the gas used \citep{tigrine2016microwave}. The chemical species present in the reactor are probed at steady-state using a quadrupole mass spectrometer (\textit{Hiden Analytical EQP 200 QMS}) to measure neutral compounds and positive ions with a high sensitivity below parts-per-million (ppm) level \citep{bourgalais2019low,dubois2019situ}. \par
In this work, we use a surfatron-type discharge with a neon gas flow that allows us to irradiate the gas mixture of the photochemical cell at an overall pressure of 0.9 mbar and at room temperature. The pressure is adjusted to represent atmospheric pressures of exoplanets but is experimentally constraint in the simulated environments. {Reaching the low pressures found in the higher layer of exoplanet environments is not possible due to experimental constraints. The configuration of the device shown in \citet{bourgalais2019low}, shows that the inlet of the MS is located very close from the outlet of the VUV lamp in order to limit wall effects and as a consequence the working pressure adjusted to ensure that reactions would take place before detection with the mass spectrometer.} The working pressure is barely higher than the pressure {(\textit{ca.} 0.01-0.1 mbar)} at which EUV radiation {penetrates} \citep{lavvas2014electron} and is low enough to avoid termolecular effects as in the actual low-density exoplanet thermospheres. Thus, the higher pressure increases the kinetics in the experiments but the results obtained on the chemistry of the ions remain transposable to the atmospheres of exoplanets. {The temperature of the thermospheres of exoplanets can reach tens of thousands of Kelvin, whereas our experiments were conducted at room temperature. However, in the lower part of the thermospheres, the deposition of stellar energy can be compensated mainly by the radiative emission of cooling ions and the subsequent ion chemistry, so that the temperature can drop to 1,000 or even a few hundred of Kelvin \citep{yelle2004aeronomy,koskinen2007thermospheric}.} {At these relatively high temperatures, the temperature dependence of the rate coefficients and branching ratios can be very important.} Therefore, it should be noted that the temperature dependence of the formation pathways of the main ions highlighted in this work, was studied with a 0D-photochemical model described in the following section. Numerical simulations performed at room temperature and at 1,000 K, show the same major ions for the two extreme cases discussed in this article. {As a conclusion, the simulated environments are relevant to study photochemical mechanisms in upper layer of cold and potentiallly warm exoplanetary atmospheres ($<$ 1,000 K) but these extrapolations at high temperatures must be taken with caution as there is little data on rate coefficients and branching ratios in the literature}. \par
A wavelength of 73.6 nm corresponding to a photon energy of 16.8 eV is used with a flux of about {10$^{14}$} ph cm$^{-2}$ s$^{-1}$ leading to an ionization rate of ca. 10$^{-6}$. This wavelength is chosen to reach the ionization threshold of the chemical species of the gas mixture that is composed of H$_2$, CO, and N$_2$. With this wavelength and the pressure in the reactor, we mimic the stellar EUV field of a star in an upper atmospheric layer \citep{france2016muscles,youngblood2016muscles,loyd2016muscles}. Transitional planets with a size between Earth and Neptune appear to have a diversity of compositions that is not well-explained but everyone agrees that low-density planets are mainly composed of molecular hydrogen but some super-Earths and mini-Neptunes will likely have thick atmospheres that are not H$_2$-dominated. Bulk atmospheric composition could have a wide variety of elemental abundance ratios in which the amount of H$_2$ has profound effects on the evolution of the atmosphere \citep{pierrehumbert2011hydrogen}. Our approach is to use gas mixtures containing H$_2$ along with other expected important molecules in upper layers of cool super-Earth atmospheres. We vary only the relative proportions of H$_2$ within a large range to test the contribution of H$_2$ in the formation of photoproducts in the cases of {H$_2$-poor} and H$_2$-rich environments which are representative of simplified super-Earth and mini-Neptune upper atmospheres \citep{rimmer2019hydrogen}. We focus on Earth-like planets by adding N$_2$ to the gas mixtures \citep{wordsworth2014abiotic} containing H$_2$ and since the bulk atmospheric composition of such environment is defined and affected largely in term of the C/O ratio, CO is adding as the only major source of carbon and oxygen to keep a ratio of 1 \citep{rimmer2019hydrogen}. Three gas mixtures have been used with different initial fractional abundances of H$_2$. The first is a gas mixture at 1\% H$_2$, the second is an intermediate at 33\% H$_2$ and finally the third is a mixture at 96\% H$_2$. The mixing ratio of N$_2$ was arbitrarily chosen equal to that of CO. Finally, water vapor is the major trace species at ppm level in the reactor coming from residual adsorption on the reactor walls.
{Contrary to a previous study on Titan conducted with the same device in \citet{bourgalais2019low}, here the reactor was not heated to minimize the presence of residual water and the pressure in the reactor is higher (0.9 mbar in this work versus 0.01 mbar in \citet{bourgalais2019low}.} \citet{gao2015stability} showed that water vapor can have profound effects on the evolution of the planet's atmosphere. Thus, the presence of water vapor traces in the reactor provides a study of the impact of trace level of water vapor in the atmosphere of exoplanets.\par
Mass spectra reported in this study are the average of 10 scans obtained at 2 s/amu with a channeltron-type detection, over the 1-50 amu mass range. In all the spectra presented below, signals with intensities lower than 1,000 cps are considered as background noise. The reported mass spectra show the direct signal recorded without background subtraction. {The duration of the irradiation is only a few minutes because, as the numerical simulations in this manuscript show, the state of equilibrium is reached in less than 1 ms for the main species because their abundance is constant. The stability of the ion signal over the 10 mass spectra recorded during the irradiation supports the claim that the steady state is reached very quickly.}\par

\subsection{0D-Photochemical Model}

{To interpret the experimental mass spectra, a coupled ion-neutral photochemical model was used to reproduce the chemistry occurring in the reactor. The model has been detailed in the literature \citep{peng2014modeling}. We provide here a short description, mentioning the main updates since the initial paper.}

\subsubsection{The Reactor Model}

{We consider a simplified 1-cell (0D) geometry of the reactor, assuming a uniform spatial distribution of the species. The gas inflow and outflow are taken into account to ensure a constant pressure, and radiative transfer is treated through Beer-Lambert-type photoabsorption, in the assumption of a uniform gas. The chemical reaction are treated as a system of ordinary differential equations (ODEs).} \par
{The implicit-explicit Runge-Kutta-Chebyshev (IRKC) method \citep{shampine2006irkc} is used  to integrate the system of partial differential equations. The photolysis and transport are treated explicitely (complex evaluation of jacobian elements for radiative-transfer makes it impractical to regard the photolysis rate equations as ordinary differential equations). The chemical ODEs are treated implicitely. The IRKC method handles stiff ODE systems \citep{verwer2004implicit}}.\par
{The fortran Reactor code is publicly available at \url{https://github.com/ppernot/Reactor}, with bash scripts for execution of parallel Monte Carlo runs on an OpenStack cloud (\href{https://github.com/ppernot/CloudReactor}{CloudReactor}). A graphical interface (\href{https://github.com/ppernot/ReactorUI}{ReactorUI}) is provided for the analysis of the results.}

\subsubsection{Chemistry}
\paragraph{The MC-ChemDB Database}
{The MC-ChemDB database (\href{https://github.com/ppernot/MC-ChemDB}{MC-ChemDB}) has been designed to handle the uncertainty on the rate parameters of chemical reactions \citep{hebrard2006photochemical,hebrard2009measurements,carrasco2008sensitivity,plessis2012production} and of photolysis processes \citep{gans2010determination,peng2014modeling}. The principle is based on a server-client architecture. The server generates and hosts random sets of databases for Monte Carlo uncertainty propagation \citep{GUMSupp1}. The client gathers the required number of samples for adapted chemical schemes: depending on the gas mixture in the reactor, only the pertinent subset of the database is used.}\par
{The database is split in three modules (neutral reactions, ionic reactions and photolysis) that require different treatments to generate random reaction rate parameters:} \\
{- Neutral chemistry (bimolecular thermal reactions and termolecular thermal recombinations) is built from \citet{hebrard2006photochemical,hebrard2009measurements,dobrijevic20161d}. Bimolecular rate constants are represented with the Kooij (modified Arrhenius) rate law, and termolecular association with a pressure-dependent Lindeman-Troe rate law \citep{dobrijevic20161d,vuitton2019simulating}. The scheme is based on partial rate constants.}\\
{- Ionic chemistry (ion-molecule reactions and dissociative recombinations) is built from \citet{carrasco2008sensitivity,plessis2012production}. Rate laws for ion-neutral reactions are mostly based on the Langevin law, with temperature-dependent laws of type ionpol \citep{wakelam2012kinetic} for molecules with non-zero dipole moments. Dissociative recombinations are based on a modified Arrhenius rate law \citep{plessis2012production}. The global reaction rates and branching ratios are treated separately.}\\
{- Photo-processes (photo-dissociation and photo-ionization), cross-sections and branching ratios (BR) are extracted from the Leiden database (\url{http://www.strw.leidenuniv.nl/~ewine/photo}) \citep{heays2017photodissociation}. If BR are not provided by Leiden, they are extracted from SWRI (\url{https://phidrates.space.swri.edu}) \citep{huebner2015photoionization}, except for CH$_4$, where the representation by \citet{peng2014modeling} is used. Cross-sections are provided with a 1 nm and 0.1 nm resolutions. At this stage, temperature-dependence of the cross-sections and BRs is ignored.}\\
{For a description of the parameters uncertainty representations, see \citet{peng2014modeling}. The database was originally oriented on the chemistry of H, C, and N-bearing species, and it has recently been upgraded to account for oxygenated species from \citet{vuitton2019simulating} and the KIDA database \citep{wakelam2012kinetic}.} 

\paragraph{The photochemical scheme}
{According to the following procedure a consistent set of reactions is iteratively generated from the irradiated initial mixture (H$_2$/N$_2$/CO/H$_2$O):\\
1. Select all the reactions involving the list of species;\\
2. Update the list of species with the generated products;\\
3. Iterate to (1) until no new species is produced.\\
The resulting chemical model contains 54 photo-processes (photolysis of N$_2$, H$_2$, CH$_4$, C$_2$H$_2$, C$_2$H$_4$, C$_2$H$_6$, HCN, NH$_3$, CO \& H$_2$O), 903 neutral reactions (811 bimolecular and 92 termolecular), 1941 ion processes (1314 bimolecular and  687 dissociative recombinations), involving 177 neutral species, 190 positive ions with masses up to 130 and electrons.}

\paragraph{Simulations}
{The model simulates chemistry under conditions similar to those of the experiments \citep{bourgalais2019low}. It is run for a time long enough to reach a stationary state (1 s) and the stationary mole fractions of the products are compared with the experimental data. Complementary Monte Carlo simulations are performed to evaluate the uncertainty on the model predictions (500 runs). A global rate analysis is performed to identify key reactions and dominant formation pathways \citep{hebrard2009measurements}.}


\subsection{Simulations of Observations with ARIEL and JWST}

We investigate the feasibility to detect the ions H$_3^+$ and H$_3$O$^+$ in sub-Neptune type planets using the radiative transfer code TauREx3 \citep{waldmann2015tau,waldmann2015rex}. Our approach consists in simulating a high resolution spectrum using TauREx in forward mode. Then we use our instrument simulators ArielRad \citep{mugnaiarielrad} to bin down the observations and estimate the instrument noise. As we are investigating the theoretical detection biases through retrieval techniques, we do not scatter the observed spectra \citep{changeat2019toward,feng2018characterizing}. Finally, we retrieve the simulated observations using TauREx3 to check whether the ion signals can be statistically recovered.  \par
We base our simulations on the bulk parameters for GJ1214b from \citet{harpsoe2013transiting}: we use a planet radius of 0.254 R$_j$, a planet mass of 0.0197 M$_j$ and a star radius of 0.216 R$_s$. In the case of ARIEL, we consider that all the Tier 3 observations for this planet are stacked together (total of 10 transits) \citep{tinetti2016science}, while for JWST we combine single observations from NIRISS and NIRSpec (total of 2 transits) to roughly match the wavelength coverage. The atmosphere is assumed to be isothermal at 700 K and filled with H$_2$ and He. We add the trace gases H$_2$O and CH$_4$ along with the two ions H$_3^+$ and H$_3$O$^+$. For our simplified model, the abundances for H$_2$O and CH$_4$ are set to 10$^{-5}$ and are constant with altitude. For the ions, we base our input abundances on profiles resembling those from \citet{helling2019lightning}. As it is expected for sub-Neptune type planets, we enhance their abundance findings by approximately one order of magnitude and use the 2-layers model \citep{changeat2019toward} to describe the abundance decrease deeper in the atmosphere. For H$_3^+$, the top abundance is 10$^{-5}$ and the surface abundance is 10$^{-14}$ with a layer pressure change at 10$^{-5}$ bar. For H$_3$O$^+$, the top abundance is also 10$^{-6}$ and the surface abundance is 10$^{-14}$. The layer change is done deeper at 10$^{-3}$ bar. On top of this, we add a grey cloud cover which is fully opaque below 5$\times$10$^{-3}$ bar as expected for this planet \citep{kreidberg2014clouds,kempton2011atmospheric,morley2013quantitatively}. {For the molecular absorptions, we use the most up-to-date molecular line lists from the ExoMol project \citep{tennyson2016exomol}, HITEMP \citep{HITEMP} and HITRAN \citep{gordon2016hitran2016} and build the cross-section tables at a resolution of 10, 000.} We also consider absorptions from Rayleigh scattering and Collision Induced Absorption (only the pairs H$_2$-H$_2$ and H$_2$-He). The list of opacities used in this paper is summarised in Table \ref{tab1}. \par

\begin{deluxetable*}{cc}
\tablenum{1}
\tablecaption{List of IR opacities used in this work.\label{tab1}}
\tablewidth{0pt}
\tablehead{
\colhead{Species} & \colhead{References}}
\startdata
H$_2$-H$_2$ & \citep{abel2011collision,fletcher2018hydrogen} \\
H$_2$-He	& \citep{abel2012infrared}\\
H$_2$O	&\citep{barton2017pressure,polyansky2018exomol}\\
CH$_4$	&\citep{hill2013temperature,yurchenko2014spectrum}\\
H$_3^+$	&\citep{mizus2017exomol}\\
H$_3$O$^+$&	Yurchenko et al. (to be published)\\
\enddata
\end{deluxetable*}

For our retrieval step, the parameter space is explored using the Multinest algorithm \citep{feroz2009use} with 500 live points and an evidence tolerance of 0.5. The list of retrieved parameters and their corresponding uniform prior bounds are shown in Table \ref{tab2}. \par

\begin{deluxetable*}{ccc}
\tablenum{2}
\tablecaption{List of retrieved parameters and their corresponding uniform prior bounds.\label{tab2}}
\tablewidth{0pt}
\tablehead{
\colhead{Parameter} & \colhead{Mode}& \colhead{Bounds}}
\startdata
Radius&Linear&0.1-0.4 R$_j$\\
H$_2$O	&Log&0.1-10$^{-12}$\\
CH$_4$	&Log&0.1-10$^{-12}$\\
H$_3^+$	&Log&0.1-10$^{-12}$\\
H$_3$O$^+$&	Log&0.1-10$^{-12}$ \\
Temperature&Linear	& 300-1500 K \\
Clouds Pressure& Log& 10-10$^{-7}$ bar \\
\enddata
\end{deluxetable*}

\section{Results}

\subsection{Identification by mass spectrometry of the neutral species formed in the experiments}

\begin{figure}[ht!]
\plotone{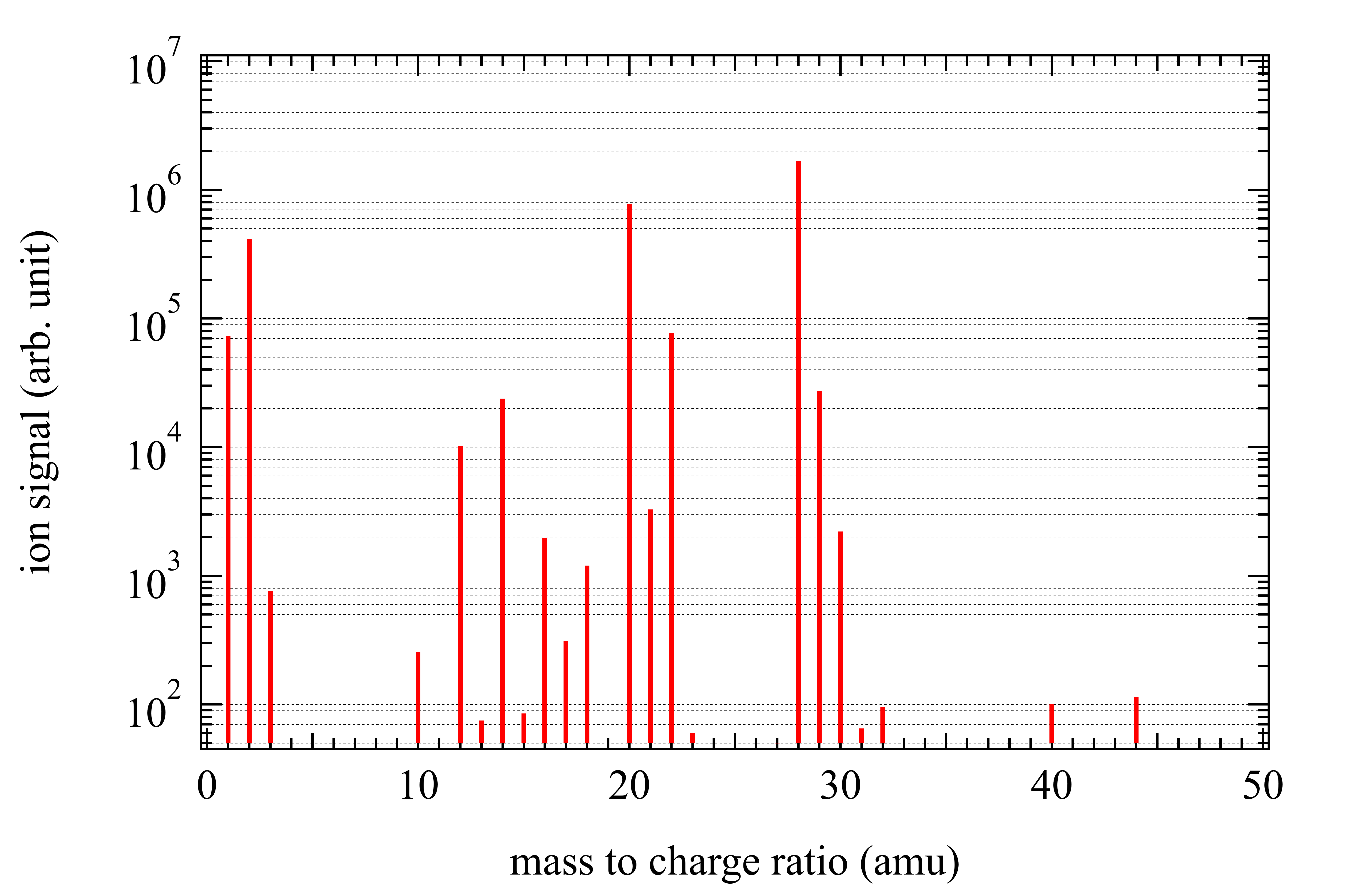}
\caption{Typical \textit{in situ} gas-phase mass spectrum of neutral species present in the reactor with a H$_2$-CO-N$_2$ gas mixture after EUV irradiation.\label{fig8}}
\end{figure}

 {A typical measured neutral mass spectrum is shown in Figure \ref{fig8} and displays the abundant peaks of the initial gas mixture species H$_2$ (\textit{m/z} = 2), CO and N$_2$ (both at \textit{m/z} = 28, with isotopologues at \textit{m/z} of 29, 30 and 31). Atomic species H, C, N and O resulting from the fragmentation of their parental molecules upon photodissociation are observed at \textit{m/z} of 1, 12, 14 and 16, respectively. Peaks at m/z of 20, 21 and 22 match the isotope relative abundances of neon, which is the non-reactive gas flowed continuously in the photochemical reactor. Peak at m/z of 10 is attributed to double ionized $^{20}$Ne$^{2+}$, produced in the mass spectrometer itself by electron impact ionization at 70 eV for the detection of neutral species. Finally, we also observed peaks at \textit{m/z} of 17 (OH) and 18 (H$_2$O).} \par
{We are conducting our experiments in an open system, where the gas mixture is introduced and pumped continuously. In the reactor, the residence time is short, making it difficult to detect neutral molecules larger than 40 amu. Neutral species beside initial gas mixture contain only atomic species H, C, N and O resulting from the fragmentation of their parental molecules upon photodissociation, neon which is the non-reactive gas owed continuously in the photochemical reactor and small molecules formed during the first reaction such as CO$_2$.} \par
{Thus, the experimental results obtained for neutral species in this work do not provide as much information and the rest of the study will focus on positive ions, about which our knowledge in this type of environment is much more limited. It should also be noted that the rate of production of ionic products relative to neutral products is similar, with molar fractions of the order of ppm relative to neutral species in the initial gas mixtures.}

\subsection{Identification by mass spectrometry of the main ions formed in the experiments}

\begin{figure}[ht!]
\plotone{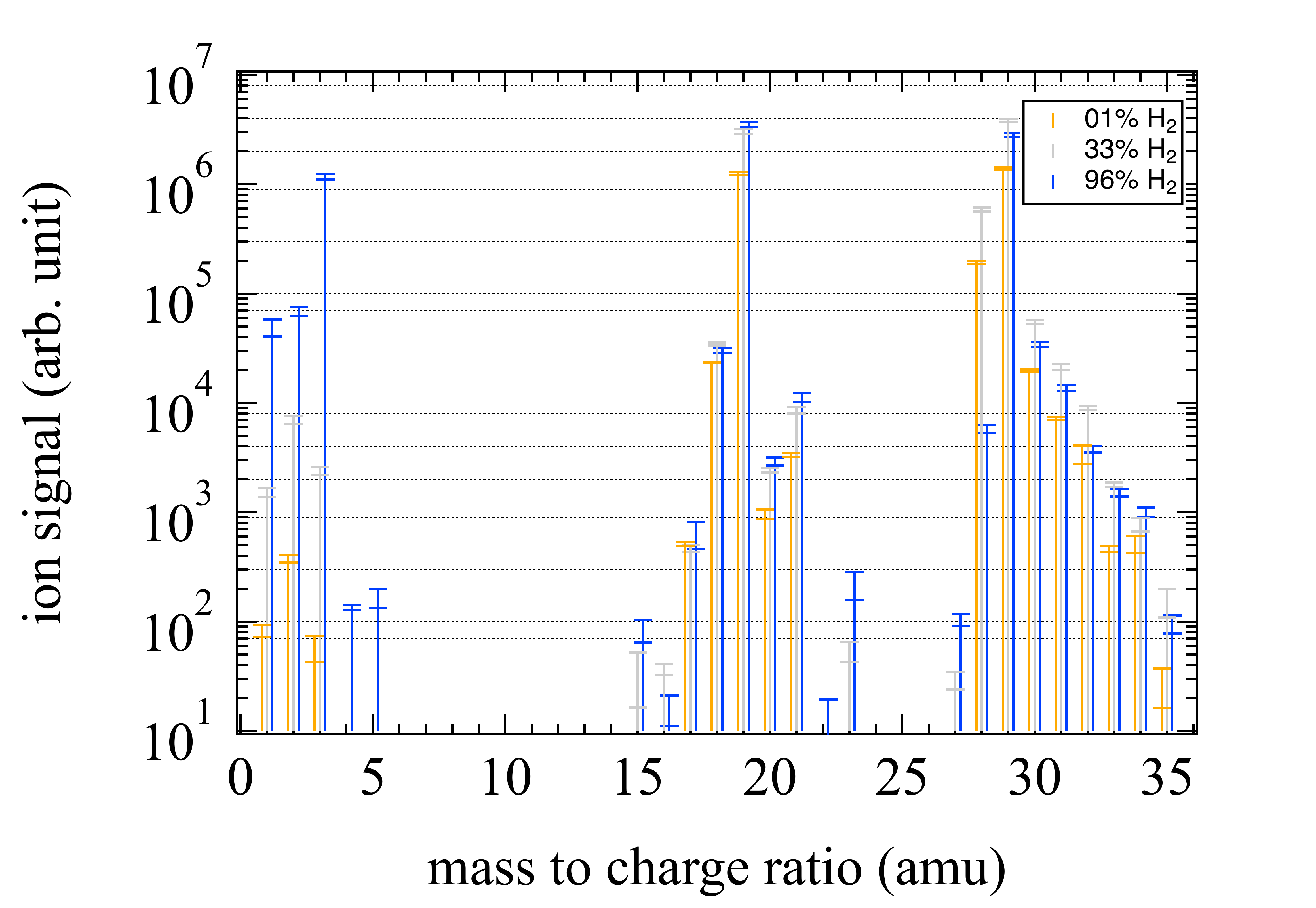}
\caption{{Positive ions} measured after irradiation at 73.6 nm of H$_2$/CO/N$_2$ gas mixtures. H$_2$ mixing ratios was chosen at 1\% (orange), 33\% (grey) and 96\% (blue) at an overall pressure in the reactor of 0.9 mbar. {Twice the standard deviation, 2$\sigma$, in 10 measurements of each mass is displayed}.\label{fig1}}
\end{figure}

Fig. \ref{fig1} shows \textit{in situ} non-normalized gas-phase mass spectra of {cationic} species in their stationary point in the reactor after 15 min EUV irradiation at an overall pressure of 0.9 mbar. The species are detected at \textit{m/z} 1, 2, 3, 18, 19, 20, 21, 28, 29, 30, 31 and 32 above the limit of the background noise. \textit{m/z} 1, 2 and 3 are attributed to H$^+$, H$_2^+$, and H$_3^+$ respectively. Due to the high sensitivity of the mass spectrometer, H$_2$O$^+$ and its protonated form H$_3$O$^+$ are detected along with first H$_3$O$^+$ isotopologues {(molecules that only differ in their isotopic composition)} at \textit{m/z} 18, 19, 20 and 21. Finally, \textit{m/z} 28, 29, 30, 31 and 32 are attributed to N$_2^+$ and CO$^+$, HCO$^+$ and/or N$_2$H$^+$ along with the first HCO$^+$ isotopologues. The first notable result is that the most abundant ions at \textit{m/z} {19} and \textit{m/z} 29 remain the same with all gas mixtures. {The formation of ions H$^+$, H$_2^+$, and H$_3^+$ ions (named H$_x$ ions) does not follow the same trend as the primary ions N$_2^+$ and CO$^+$ at mass 28 which are formed by the ionization of N$_2$ and CO. In the case of H$_2$-rich gas mixtures, H$_x$ ions, and in particular H$_3$O$^+$, dominate while the primary ions N$_2^+$ and CO$^+$ occur in significant amounts in the case of gas mixtures where H$_2$ is not in majority}.

\subsection{Unraveling the chemical processes through ion-neutral modelling}
\subsubsection{Hydrogen-poor atmospheres}

\begin{figure}[ht!]
\plotone{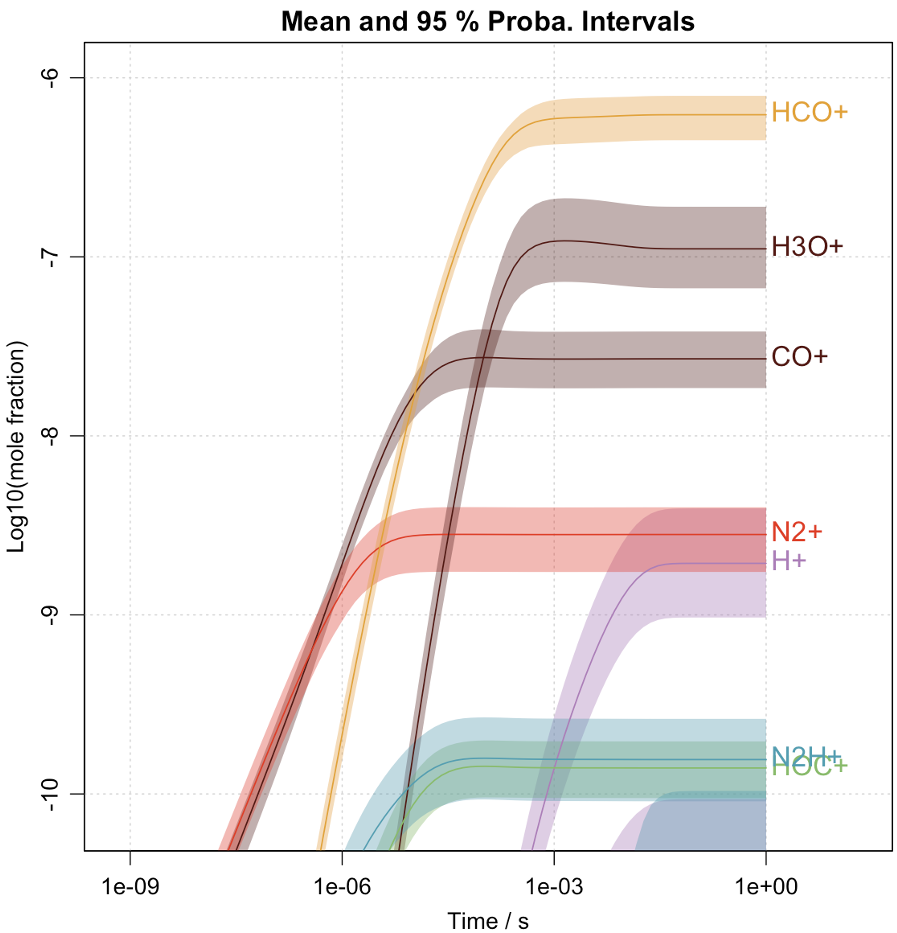}
\caption{Simulated evolution of the mole fractions of the main ionic photo-products after irradiation at 73.6 nm of H$_2$/CO/N$_2$ gas mixture with a H$_2$ mixing ratio of 1\%. All species with a mole fraction above 10$^{-11}$ are represented.\label{fig2}}
\end{figure}

Fig. \ref{fig2} displays the simulated evolution of the mole fractions of the main ionic photo-products after irradiation at 73.6 nm of H$_2$/CO/N$_2$ gas mixture with a H$_2$ mixing ratio of 1\%. The two most abundant predicted ions using the 0D-photochemical model are HCO$^+$ and H$_3$O$^+$ as observed in the experimental mass spectrum of Fig. \ref{fig1}. The main HCO$^+$ formation pathways are found to be through the reactions R1, R2, and R3. HOC$^+$ is formed in a second exit channel of R1 but is rapidly converted into HCO$^+$ through proton transfer (R3).

\begin{reactempheq}[]{align}
\ce{CO+ + H$_2$ &-> HCO$^+$/HOC$^+$ + H} \\
\ce{N$_2$H+ + CO &-> HCO$^+$ + N$_2$}\\
\ce{HOC+ + CO &-> HCO$^+$ + CO}
\end{reactempheq}

The main loss pathways of HCO$^+$ are found to be dissociative recombination with electrons, along with the proton transfer reaction with water (R4 and R5):

\begin{reactempheq}{align}
\ce{HCO$^+$ + e$^-$ &-> H + CO}\\
\ce{HCO$^+$ + H$_2$O &-> H$_3$O$^+$ + CO}
\end{reactempheq}

R5 is the only formation pathway for H$_3$O$^+$, using trace water vapor in the reactor. H$_3$O$^+$ is consumed through dissociative recombination leading to water and OH radicals. Modeling of chemistry within the reactor also shows that the formation of N$_2$H$^+$ is negligible compared to that of HCO$^+$. Its main production pathways are:

\begin{reactempheq}{align}
\ce{HOC+ + N$_2$ &-> N$_2$H+ + CO}\\
\ce{N$_2$+ + H$_2$ &-> N$_2$H+ + H}
\end{reactempheq}

The low abundance of N$_2$H$^+$ compared to HCO$^+$ is due to the efficient reaction R2 showing the propensity of CO to consume N$_2$H$^+$ to produce HCO$^+$. As observed in the experimental mass spectrum, {the H$_x$ ions are expected to be negligible through the main reactions (R8, R9, and R10)}. 

\begin{reactempheq}{align}
\ce{CO+ + H &-> H$^+$ + CO}\\
\ce{H$_2$ + h$\nu$ &-> H$_2^+$ + e$^-$ }\\
\ce{HOC+ + H$_2$ &-> H$_3^+$ + CO}
\end{reactempheq}

HOC$^+$ indeed mainly reacts with N$_2$ and CO (R3 and R6) to produce N$_2$H$^+$ and HCO$^+$ respectively rather than with H$_2$.

\subsubsection{Hydrogen-rich atmospheres}

\begin{figure}[ht!]
\plotone{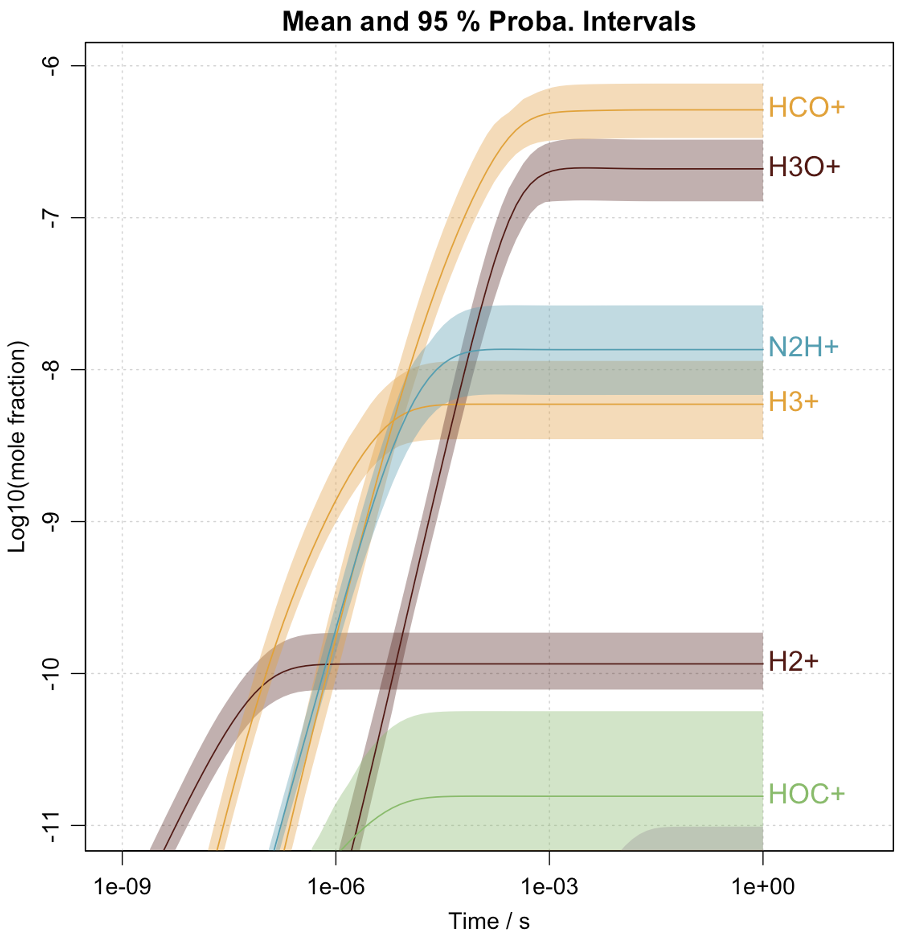}
\caption{Simulated evolution of the mole fractions of the main ionic photo-products after irradiation at 73.6 nm of H$_2$/CO/N$_2$ gas mixture with a H$_2$ mixing ratio of 96\%. All species with a mole fraction above 10$^{-11}$ are represented.\label{fig3}}
\end{figure}

Fig. \ref{fig3} displays the simulated evolution of the mole fractions of the main ionic photo-products after irradiation at 73.6 nm of H$_2$/CO/N$_2$ gas mixture with a H$_2$ mixing ratio of 96\%. The two most abundant predicted ions using the 0D-photochemical model remain HCO$^+$ and H$_3$O$^+$ as observed in the experimental mass spectrum of Fig. \ref{fig1}. However, the main formation pathways for HCO$^+$ is still via R3 but with an additional parallel reaction involving H$_3^+$:

\begin{reactempheq}{align}
\ce{H$_3^+$ + CO &-> HCO$^+$ + H$_2$}
\end{reactempheq}

Independent of the mixing ratio of H$_2$, the main destruction pathways of HCO$^+$ remain the same in both H$_2$ environments (R4 and R5) which includes the main production pathway for H$_3$O$^+$ (R5). Compared to the case of H$_2$-poor atmospheres, the mole fraction of N$_2$H$^+$ increases significantly in the presence of a large abundance of H$_2$ due to a production pathway more efficient than R6 and R7:

\begin{reactempheq}{align}
\ce{H$_3^+$ + N$_2$ &-> N$_2$H+ + H$_2$}
\end{reactempheq}

In this H$_2$-rich environment, H$_3^+$ starts to be the dominant ion due to a more efficient formation pathway:

\begin{reactempheq}{align}
\ce{H$_2$+ + H$_2$ &-> H$_3^+$ + H}
\end{reactempheq}

which supersedes R10. Once formed H$_3^+$ leads to the formation of the dominant ions HCO$^+$ and N$_2$H$^+$ by R11 and R12 respectively.

\subsubsection{Comparison of the two cases}

Laboratory simulations show that the same main ions (HCO$^+$, H$_3$O$^+$) are observed in H$_2$- poor and rich environments. This implies that we can expect to observe similar contributions of these ions in the thermosphere of super-Earths and mini-Neptunes. However, they are formed through different chemical pathways by H-transfer reactions leading to the observed stable protonated ions. Another remarkable result is that even with trace amounts of water, the formation of the hydronium ion (H$_3$O$^+$) is very effective. These results highlight the need to accurately model ion-neutral chemistry in atmospheric models of exoplanets. In the case of the low H$_2$ environment, the primary ions N$_2^+$ and CO$^+$ formed at 73.6 nm trigger the molecular growth reacting with H$_2$. In the specific case of environments rich in H$_2$, the triatomic hydrogen ion H$_3^+$ is one of the most abundant ions and is shown to be very reactive, leading to the formation of heavier ions (N$_2$H$^+$ and HCO$^+$). 

\subsection{Simulations of Observations with ARIEL and JWST
}
\subsubsection{Choice of the planet type and the ion targets
}

The fact that the H$_3^+$ ion is predicted to be abundant and highly reactive leading to the formation of heavier ions in H$_2$-rich environments is of great interest in the context of exoplanetary observations considering the upcoming launches of the ARIEL and JWST telescopes. The abundance and distribution of H$_3^+$, the stable ionic form of H$_2$ is essential to understand the chemistry and to get information about the thermal structure, the dynamic, and the energy balance of exoplanet atmospheres \citep{miller2000infrared}. H$_3^+$ has been thoroughly studied in the mid-InfraRed (IR) on Jupiter \citep{stallard2001dynamics}, Saturn \citep{geballe1993detection} and Uranus \citep{trafton1993detection} but detection in the thermosphere of hot Jupiter exoplanets, with a predicted mixing ratio ranging from 10$^{-6}$ up to 10$^{-4}$, so far remains illusive \citep{lenz2016crires,shkolnik2006no}. \citet{helling2019lightning} assume that the formation pathway of H$_3^+$ through the ionization of H$_2$ (R11), although very efficient, is difficult in hot exoplanet atmospheres due to the thermal decomposition of H$_2$ which becomes important when temperature reaches 1,000 K \citep{yelle2004aeronomy}. Indeed, the very hot thermosphere of these exoplanets ($>$ 8,000 K) combined with very intense stellar irradiations promotes the formation of atomic neutral and H$^+$ making the detection of H$_3^+$ difficult \citep{koskinen2010characterizing}. However, H$_3^+$ is an efficient coolant up to 10,000 K, temperature above which thermal dissociation becomes significant and this thermostat effect is important in controlling the atmospheric stability of giant exoplanets \citep{neale1996spectroscopic,miller2010h,miller2013cooling}. By lowering the temperature of the exosphere and the related pressure, H$_3^+$ helps to counteract the atmospheric escape of atomic and molecular species. Theory predicts that the cooling potential of H$_3^+$ would allow one to observe H$_3^+$ in giant exoplanets inside 0.2 {to} 1 AU orbits \citep{yelle2004aeronomy,miller2013cooling,koskinen2007thermospheric}. H$_3^+$ is able to offset the increased heating due to EUV radiation which allows to produce increased ion densities by photoionization. Thus, to get sufficient H$_3^+$ an atmosphere relatively cold or far-enough from its star to have a low-temperature thermosphere is needed. The results of the laboratory experiments in this work suggest that colder atmospheres dominated by H$_2$ would be more conducive to H$_2$ stability leading to a higher H$_3^+$ abundance. Warm Neptunes (ca. 400 {to} 800 K) would therefore potentially be better candidates than hot Jupiters for the detection of H$_3^+$. \\
{Like H$_3^+$, long lifetimes associated with certain excited states can lead to population trapping and unexpected state distributions in non-thermalized environments. A recent theoretical study by \citet{melnikov2016radiative} calculated the stability of the ro-vibrational states of H$_3$O$^+$, the lifetimes of individual states and the overall cooling rates. H$_3$O$^+$ is present in abundance in diffuse and dense regions of the interstellar medium, such as comets and molecular clouds \citep{goicoechea2001far,barber2009water}. So far, no attempt has been made to observe H$_3$O$^+$ in the thermosphere of exoplanets, although it may, like H$_3^+$, act as a cooling agent and bring some constraints on the physical parameters of the atmosphere.}

\subsubsection{Simulation of the observation of ions in a GJ1214b-like planet by ARIEL and JWST}

\begin{figure}[ht!]
\plotone{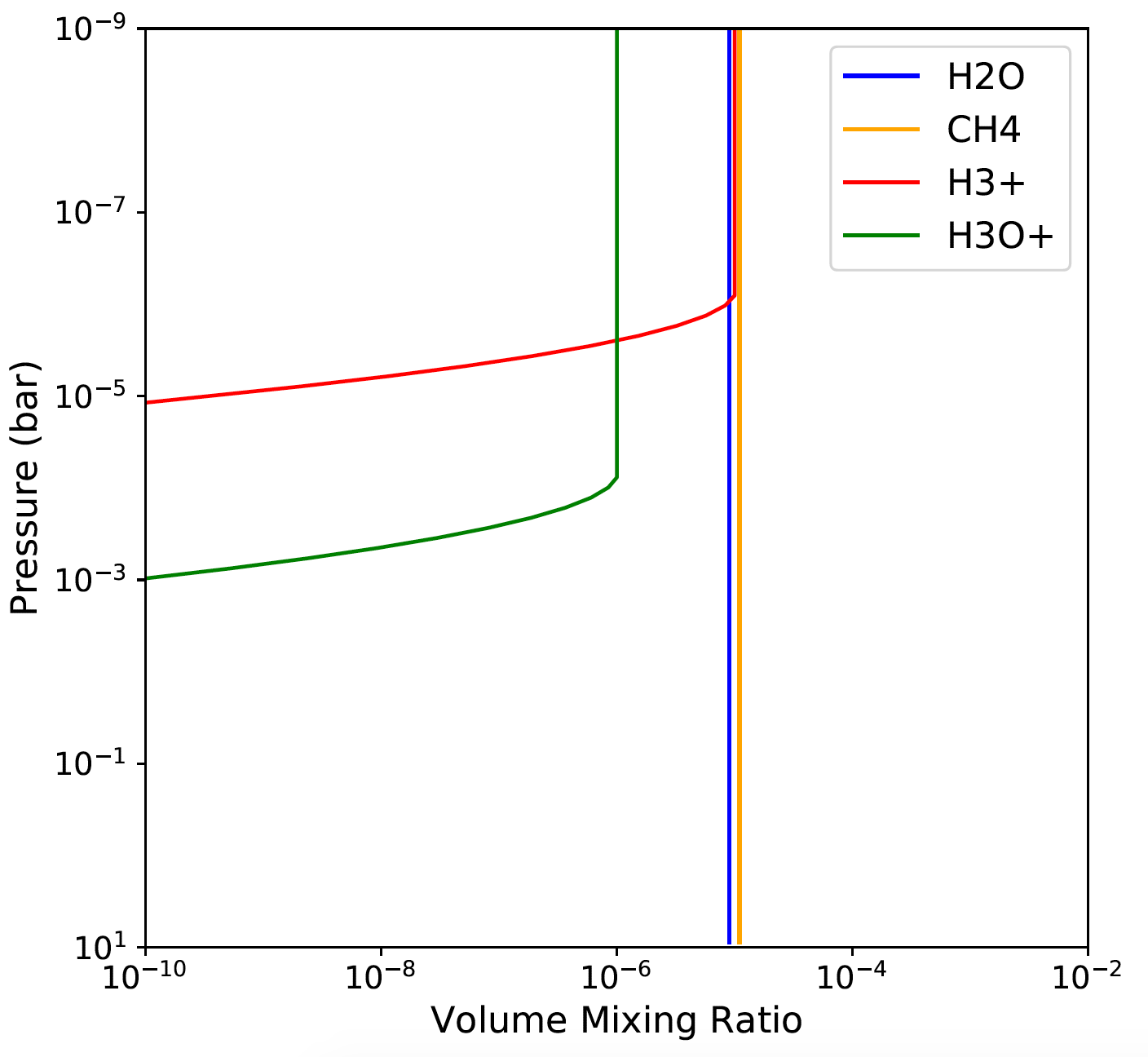}
\caption{Atmospheric profiles used as input for foward simulations of exoplanet spectrum, as well as for fitting data in retrievals.\label{fig7}}
\end{figure}

In order to evaluate the ability of future telescopes to detect these two ions, we simulate a simplified atmosphere for a GJ1214b-type warm Neptune planet as observed by JWST and ARIEL with respectively 2 and 10 stacked transits. The chemical profiles we use for the H$_3^+$ and H$_3$O$^+$ ions are inspired by the work of \citet{helling2019lightning}; we also include the absorption of H$_2$O and CH$_4$, which are the main absorbers. 
{For our simplified model, the abundances for water vapor and methane are set to 10 ppm and are constant with altitude. Those two molecules are predicted to be abundant in sub-Neptune planets \citep{kempton2011atmospheric,venot2014atmospheric,hu2015photochemistry,tinetti2018chemical,changeat2019toward} and have very strong features in the wavelength range covered by Ariel and JWST. {As there are still few models of exoplanets to incorporate reactions with ions, the abundances of H$_3^+$ and H$_3$O$^+$ that can be expected remain unclear. We present the results with realistic mixing ratios inspired by the work of \citet{helling2019lightning} though. The purpose of these simulations was to determine with the resolution of future telescopes if these mixing ratios would allow the cations to be detected in the atmospheres of exoplanets and through which transitions.} While their profiles may variate with altitude, for the purpose of our simplified example we consider that they are vertically mixed, hence constant with altitude (see Fig.\ref{fig7})}.\par

{It should be noted that no object exactly similar to our experimental gas mixtures has yet been identified, but the fact that CH$_4$ replaces CO/N$_2$ among the secondary species does not change the final result, which is that H$_3$O$^+$ and H$_3^+$ will be among the major ions formed in a H$_2$-dominated environment where there is the presence of oxygen, even if only in trace amounts \citep{hollenbach2012chemistry,beuther2014protostars,gerin2016interstellar}. Efficient series of proton transfer reactions from H$_3^+$ or other abundant ions, depending on the secondary species in the environments, will easily lead to the formation of H$_3$O$^+$ \citep{indriolo2013cosmic,van2013interstellar}, making the detection of H$_3$O$^+$ and H$_3^+$ in an object like GJ1214b quite legitimate.}\par
\begin{figure}[ht!]
\plotone{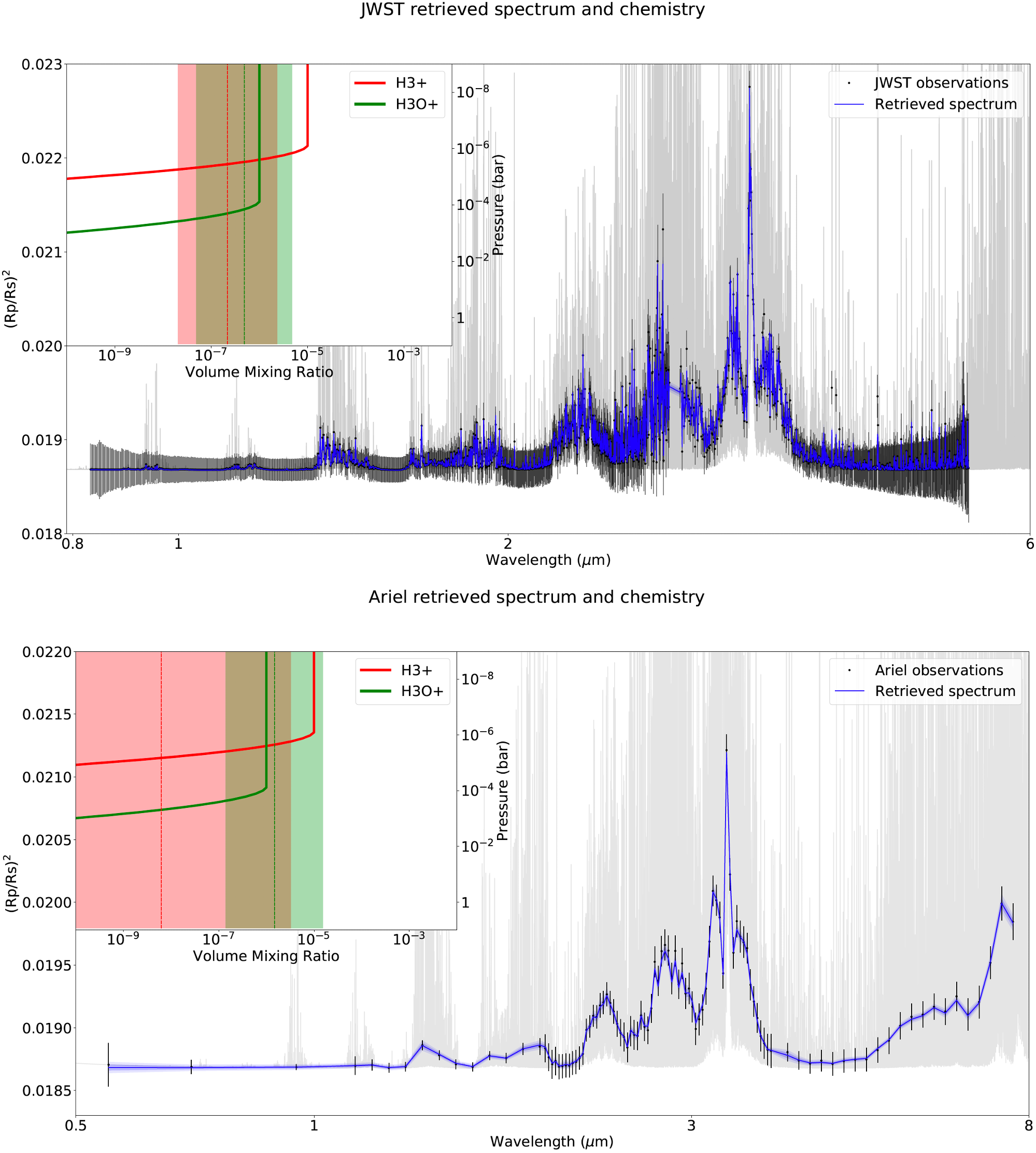}
\caption{Retrieved spectra and volume mixing ratios for our JWST (top) and ARIEL (bottom) simulations of a GJ1214b-like planet with H$_3$O$^+$ and H$_3^+$ ions. For the abundances the solid line is the true value, the dashed line the mean retrieved value and the shaded area represents the 1$\sigma$ retrieved value. \label{fig4}}
\end{figure}
\begin{figure}[ht!]
\plotone{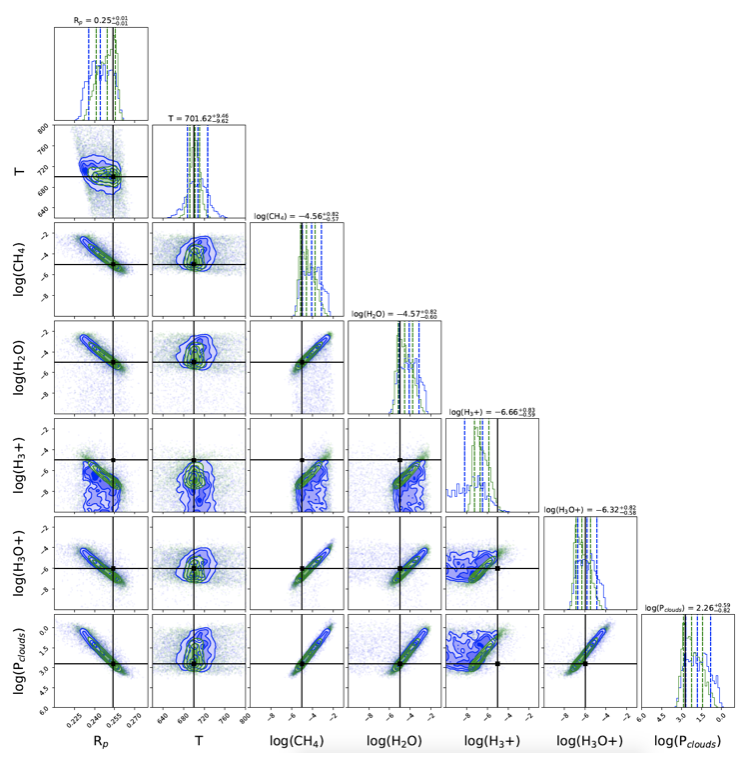}
\caption{Posterior distributions for ARIEL (blue) and JWST (green) retrievals of our GJ1214b planet case.\label{fig5}}
\end{figure}
\begin{figure}[ht!]
\plotone{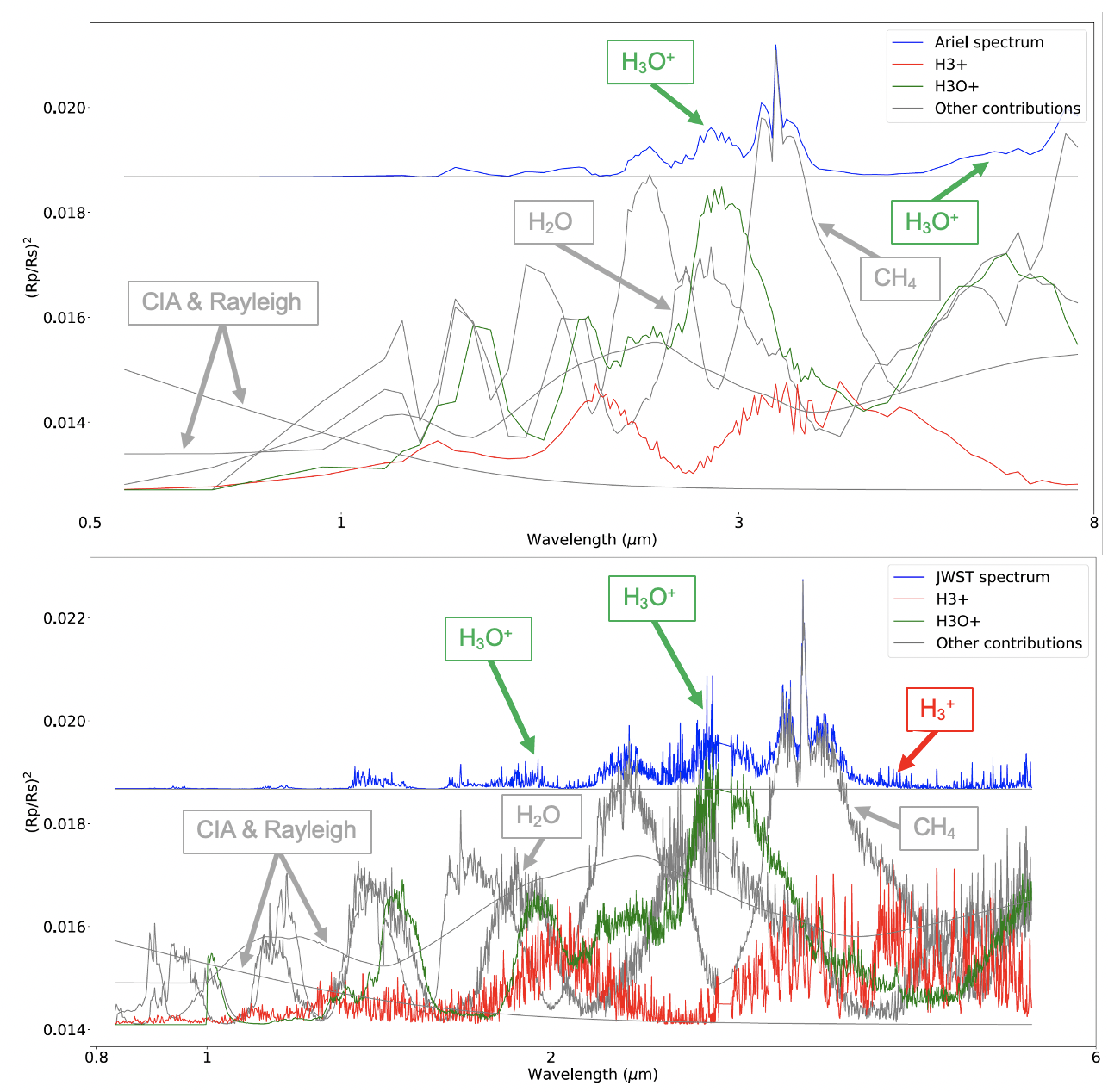}
\caption{Contribution of the different absorbers to the transmission spectra of our best fit retrievals for ARIEL (top) and JWST (bottom). \label{fig6}}
\end{figure}

In both ARIEL and JWST cases, our retrieval analysis manages to reproduce the observed spectra (see Fig. \ref{fig4}). We can see that H$_3$O$^+$ is detectable by both instruments (see also posterior distributions Fig. \ref{fig5}). Indeed, H$_3$O$^+$ presents a wide feature around 2.8 $\mu$m, which is easily captured. H$_3^+$, however, exhibits much weaker features that are largely masked by the clouds. In general, the grey cloud cover, which represents a pessimistic assumption, introduces a large degeneracy with the planet radius and the retrieved abundances (see Fig. \ref{fig6}). This, along with the assumption of constant chemistry in the retrievals, explain the slight difference from the true and retrieved chemistry for H$_2$O and CH$_4$. H$_3^+$ is only successfully detected in the JWST simulated spectrum thanks to the presence of small features, only seen at high resolution. In the ARIEL case, an upper limit of 10$^{-5}$ on H$_3^+$ abundance can be deduced but the detection remains unclear. In order to detect this ion with ARIEL, higher abundances or a slightly higher number of transits will be necessary.

\section{Conclusion}
Laboratory experiments simulating thermospheric chemistry in simplified atmospheres of super-Earth and mini-Neptunes exoplanets were performed. A photochemical reactor containing gas mixtures with different mixing ratios of H$_2$ and important exoplanetary trace compounds (CO and N$_2$) were irradiated at a pressure of 0.9 mbar, representative of high altitude in an exoplanetary atmosphere, and a wavelength of 73.6 nm. The photoproduced ions were measured using a quadrupole mass spectrometer and the experimental data were compared to the expected abundances of a 0D-photochemical model. A sensitivity analysis based on the Monte Carlo method allows us to confirm H$_3^+$ as one of the main ions in environments rich in H$_2$. Our retrieval analysis based on these simulations show that in realistic conditions, the next generation of space telescopes should be able to accurately constrain H$_3$O$^+$ abundances in sub-Neptunes due to the large broadband feature at around 2.8 $\mu$m. For H$_3^+$, the strongest features seem to require high resolution, which can easily be achieved by JWST but would require the stacking of a large number of observations in the case of ARIEL. Even if the presence of other molecules make detections more difficult, spectral retrieval models provide a powerful technique to disentangle the contribution of the different species from the data \citep{lammer2019role,lavie2017helios,goyal2017library,gandhi2017retrieval}. Additionally, H$_3^+$ is very reactive and tends to donate a proton to all the molecules it encounters, enriching chemistry through chain reactions to form more complex and diverse molecules. H$_3^+$ will thus be difficult to observe if it reacts very quickly with neighboring molecules such as CO and N$_2$, as shown in this work. Given the results of this work, we suggest that it should be fruitful to search upper exoplanetary atmospheres for abundant protonated ions such as H$_3$O$^+$ produced by reactions with H$_3^+$. The simultaneous detection of H$_3^+$ and H$_3$O$^+$ ions would serve to constrain parameters when characterizing exoplanets.\par
Several future ground-based and space telescopes (JWST, ARIEL, E-ELT, TMT) will allow the characterization of the atmospheres of exoplanets using infrared spectroscopy, in particular to search for biosignatures within the atmospheres of rocky planets. Cool stars such as M dwarfs are preferred targets for the search of rocky temperate planets, as they have a closer habitable zone due to their size and temperature, which favors the detection and characterization of terrestrial exoplanets \citep{tarter2007reappraisal,scalo2007m}. However, the exposure of planetary atmospheres to the strong UV radiation from these active stars may have drastically negative effect on the habitability of planets \citep{luger2015extreme,tian2009thermal,erkaev2013xuv,miguel2014effect,chadney2016euv}. High-energy radiation drives the chemical composition, including the formation of IR-radiating species that will influence the evolution of the atmosphere by controlling its thermal structure. Species like H$_3^+$ and H$_3$O$^+$ with different cooling rates will allow different EUV-response to thermospheric expansion and hinder the escape of atoms and other volatiles that could be required for life. Thus, the detection of these ions can provide information on the physical parameters of atmospheres and the formation of atmospheric constituents. {For instance the saturated molecular hydronium ion is known to play an important role in planetary and interstellar oxygen chemistry network in a way that dissociative recombination of hydronium ions with electrons could be a source of target molecules related to habitability like water vapor and O$_2$ \citep{larsson2008dissociative}. Thus H$_3^+$ or H$_3$O$^+$ may be used to estimate water vapor and O$_2$ abundances when direct detection is unfeasible.} \par
As a conclusion, this work proposed H$_3^+$ and H$_3$O$^+$ as potential tracers of the nature and evolution of low-gravity exoplanets. H$_3^+$ formation is negligible in low H$_2$ environments but important in H$_2$-rich gas mixtures, while H$_3$O$^+$ should remain abundant in both environments if water vapor is present in the atmosphere even at trace level. Thus, non-detection of H$_3^+$ associated with the observation of H$_3$O$^+$ in an exoplanetary atmosphere would provide additional information for the classification of planets in the transition between super-Earths and mini-Neptunes. \par
Future studies should focus on the potential detection of the other major ions highlighted in this work, HCO$^+$ and N$_2$H$^+$, which may be relatively abundant within the atmospheres of exoplanets where CO and N$_2$ would be present. However, theoretical calculations are required to provide a list of absorption transitions and their related intensities for these species. One strategy to support the theoretical calculations would be to develop experimental devices to measure ion absorption cross sections in IR. {Experimental devices must begin to focus on the evolution of thermospheric chemistry as a function of temperature, which requires the development of simulation chambers that can be temperature-controlled between 300 and 1,000 K.} Finally, photochemical models of exoplanets should begin to take into account the coupling between the chemistry of neutral species and ions for the simulation of atmospheres of cold and warm exoplanets.\par

\acknowledgments
This research was supported by the European Research Council Starting Grant PRIMCHEM 636829 to N.C.. O.V. thanks the CNRS/INSU Programme National de Plan\'etologie (PNP) and the CNES for funding support. We acknowledge the help and the fruitful discussions with Dr Billy Edwards, who provided access to his JWST noise simulator ExoWebb. 

\bibliography{sample63}{}

\begin{thebibliography}{}
\expandafter\ifx\csname natexlab\endcsname\relax\def\natexlab#1{#1}\fi
\providecommand{\url}[1]{\href{#1}{#1}}
\providecommand{\dodoi}[1]{doi:~\href{http://doi.org/#1}{\nolinkurl{#1}}}
\providecommand{\doeprint}[1]{\href{http://ascl.net/#1}{\nolinkurl{http://ascl.net/#1}}}
\providecommand{\doarXiv}[1]{\href{https://arxiv.org/abs/#1}{\nolinkurl{https://arxiv.org/abs/#1}}}

\bibitem[{Abel {et~al.}(2011)Abel, Frommhold, Li, \& Hunt}]{abel2011collision}
Abel, M., Frommhold, L., Li, X., \& Hunt, K.~L. 2011, The Journal of Physical
  Chemistry A, 115, 6805

\bibitem[{Abel {et~al.}(2012)Abel, Frommhold, Li, \& Hunt}]{abel2012infrared}
---. 2012, The Journal of chemical physics, 136, 044319

\bibitem[{Barber {et~al.}(2009)Barber, Miller, Dello~Russo, Mumma, Tennyson, \&
  Guio}]{barber2009water}
Barber, R.~J., Miller, S., Dello~Russo, N., {et~al.} 2009, Monthly Notices of
  the Royal Astronomical Society, 398, 1593

\bibitem[{Barton {et~al.}(2017)Barton, Hill, Yurchenko, Tennyson, Dudaryonok,
  \& Lavrentieva}]{barton2017pressure}
Barton, E.~J., Hill, C., Yurchenko, S.~N., {et~al.} 2017, Journal of
  Quantitative Spectroscopy and Radiative Transfer, 187, 453

\bibitem[{Beuther {et~al.}(2014)Beuther, Klessen, Dullemond, \&
  Henning}]{beuther2014protostars}
Beuther, H., Klessen, R.~S., Dullemond, C.~P., \& Henning, T.~K. 2014,
  Protostars and planets VI (University of Arizona Press)

\bibitem[{{BIPM} {et~al.}(2008){BIPM}, {IEC}, {IFCC}, {ILAC}, {ISO}, {IUPAC},
  {IUPAP}, \& {OIML}}]{GUMSupp1}
{BIPM}, {IEC}, {IFCC}, {et~al.} 2008, {E}valuation of measurement data -
  {S}upplement 1 to the "{G}uide to the expression of uncertainty in
  measurement" - {P}ropagation of distributions using a {Monte Carlo} method,
  Tech. Rep. 101:2008, Joint Committee for Guides in Metrology, JCGM.
\newblock
  \url{http://www.bipm.org/utils/common/documents/jcgm/JCGM_101_2008_E.pdf}

\bibitem[{Bolmont {et~al.}(2016)Bolmont, Selsis, Owen, Ribas, Raymond, Leconte,
  \& Gillon}]{bolmont2016water}
Bolmont, E., Selsis, F., Owen, J.~E., {et~al.} 2016, Monthly Notices of the
  Royal Astronomical Society, 464, 3728

\bibitem[{{Borucki}(2016)}]{2016RPPh...79c6901B}
{Borucki}, W.~J. 2016, Reports on Progress in Physics, 79, 036901,
  \dodoi{10.1088/0034-4885/79/3/036901}

\bibitem[{Bourgalais {et~al.}(2019)Bourgalais, Carrasco, Vettier, \&
  Pernot}]{bourgalais2019low}
Bourgalais, J., Carrasco, N., Vettier, L., \& Pernot, P. 2019, Journal of
  Geophysical Research: Space Physics, 124, 9214

\bibitem[{Bourrier {et~al.}(2017)Bourrier, Ehrenreich, Wheatley, Bolmont,
  Gillon, de~Wit, Burgasser, Jehin, Queloz, \&
  Triaud}]{bourrier2017reconnaissance}
Bourrier, V., Ehrenreich, D., Wheatley, P., {et~al.} 2017, Astronomy \&
  Astrophysics, 599, L3

\bibitem[{Carrasco {et~al.}(2008)Carrasco, Alcaraz, Dutuit, Plessis, Thissen,
  Vuitton, Yelle, \& Pernot}]{carrasco2008sensitivity}
Carrasco, N., Alcaraz, C., Dutuit, O., {et~al.} 2008, Planetary and Space
  Science, 56, 1644

\bibitem[{Carrasco {et~al.}(2013)Carrasco, Giuliani, Correia, \&
  Cernogora}]{carrasco2013vuv}
Carrasco, N., Giuliani, A., Correia, J.-J., \& Cernogora, G. 2013, Journal of
  synchrotron radiation, 20, 587

\bibitem[{Chabrier(2003)}]{chabrier2003galactic}
Chabrier, G. 2003, Publications of the Astronomical Society of the Pacific,
  115, 763

\bibitem[{Chadney {et~al.}(2016)Chadney, Galand, Koskinen, Miller,
  Sanz-Forcada, Unruh, \& Yelle}]{chadney2016euv}
Chadney, J., Galand, M., Koskinen, T., {et~al.} 2016, Astronomy \&
  Astrophysics, 587, A87

\bibitem[{Chadney {et~al.}(2015)Chadney, Galand, Unruh, Koskinen, \&
  Sanz-Forcada}]{chadney2015xuv}
Chadney, J., Galand, M., Unruh, Y., Koskinen, T., \& Sanz-Forcada, J. 2015,
  Icarus, 250, 357

\bibitem[{Changeat {et~al.}(2019)Changeat, Edwards, Waldmann, \&
  Tinetti}]{changeat2019toward}
Changeat, Q., Edwards, B., Waldmann, I., \& Tinetti, G. 2019, The Astrophysical
  Journal, 886, 39

\bibitem[{Coughlin {et~al.}(2016)Coughlin, Mullally, Thompson, Rowe, Burke,
  Latham, Batalha, Ofir, Quarles, Henze, {et~al.}}]{coughlin2016planetary}
Coughlin, J.~L., Mullally, F., Thompson, S.~E., {et~al.} 2016, The
  Astrophysical Journal Supplement Series, 224, 12

\bibitem[{Dobrijevic {et~al.}(2016)Dobrijevic, Loison, Hickson, \&
  Gronoff}]{dobrijevic20161d}
Dobrijevic, M., Loison, J., Hickson, K., \& Gronoff, G. 2016, Icarus, 268, 313

\bibitem[{Drummond {et~al.}(2016)Drummond, Tremblin, Baraffe, Amundsen, Mayne,
  Venot, \& Goyal}]{drummond2016effects}
Drummond, B., Tremblin, P., Baraffe, I., {et~al.} 2016, Astronomy \&
  Astrophysics, 594, A69

\bibitem[{Dubois {et~al.}(2019)Dubois, Carrasco, Petrucciani, Vettier, Tigrine,
  \& Pernot}]{dubois2019situ}
Dubois, D., Carrasco, N., Petrucciani, M., {et~al.} 2019, Icarus, 317, 182

\bibitem[{Elkins-Tanton \& Seager(2008)}]{elkins2008ranges}
Elkins-Tanton, L.~T., \& Seager, S. 2008, The Astrophysical Journal, 685, 1237

\bibitem[{Erkaev {et~al.}(2013)Erkaev, Lammer, Odert, Kulikov, Kislyakova,
  Khodachenko, G{\"u}del, Hanslmeier, \& Biernat}]{erkaev2013xuv}
Erkaev, N.~V., Lammer, H., Odert, P., {et~al.} 2013, Astrobiology, 13, 1011

\bibitem[{Feng {et~al.}(2018)Feng, Robinson, Fortney, Lupu, Marley, Lewis,
  Macintosh, \& Line}]{feng2018characterizing}
Feng, Y.~K., Robinson, T.~D., Fortney, J.~J., {et~al.} 2018, The Astronomical
  Journal, 155, 200

\bibitem[{Feroz {et~al.}(2009)Feroz, Gair, Hobson, \& Porter}]{feroz2009use}
Feroz, F., Gair, J.~R., Hobson, M.~P., \& Porter, E.~K. 2009, Classical and
  Quantum Gravity, 26, 215003

\bibitem[{Fletcher {et~al.}(2018)Fletcher, Gustafsson, \&
  Orton}]{fletcher2018hydrogen}
Fletcher, L.~N., Gustafsson, M., \& Orton, G.~S. 2018, The Astrophysical
  Journal Supplement Series, 235, 24

\bibitem[{France {et~al.}(2016)France, Loyd, Youngblood, Brown, Schneider,
  Hawley, Froning, Linsky, Roberge, Buccino, {et~al.}}]{france2016muscles}
France, K., Loyd, R.~P., Youngblood, A., {et~al.} 2016, The Astrophysical
  Journal, 820, 89

\bibitem[{Gandhi \& Madhusudhan(2017)}]{gandhi2017retrieval}
Gandhi, S., \& Madhusudhan, N. 2017, Monthly Notices of the Royal Astronomical
  Society, 474, 271

\bibitem[{Gans {et~al.}(2010)Gans, Mendes, Boy{\'e}-P{\'e}ronne, Douin, Garcia,
  Soldi-Lose, de~Miranda, Alcaraz, Carrasco, Pernot,
  {et~al.}}]{gans2010determination}
Gans, B., Mendes, L. A.~V., Boy{\'e}-P{\'e}ronne, S., {et~al.} 2010, The
  Journal of Physical Chemistry A, 114, 3237

\bibitem[{Gao {et~al.}(2015)Gao, Hu, Robinson, Li, \& Yung}]{gao2015stability}
Gao, P., Hu, R., Robinson, T.~D., Li, C., \& Yung, Y.~L. 2015, The
  Astrophysical Journal, 806, 249

\bibitem[{Gardner {et~al.}(2006)Gardner, Mather, Clampin, Doyon, Greenhouse,
  Hammel, Hutchings, Jakobsen, Lilly, Long, {et~al.}}]{gardner2006james}
Gardner, J.~P., Mather, J.~C., Clampin, M., {et~al.} 2006, Space Science
  Reviews, 123, 485

\bibitem[{Geballe {et~al.}(1993)Geballe, Jagod, \& Oka}]{geballe1993detection}
Geballe, T., Jagod, M.-F., \& Oka, T. 1993, The Astrophysical Journal, 408,
  L109

\bibitem[{Gerin {et~al.}(2016)Gerin, Neufeld, \&
  Goicoechea}]{gerin2016interstellar}
Gerin, M., Neufeld, D.~A., \& Goicoechea, J.~R. 2016, Annual Review of
  Astronomy and Astrophysics, 54, 181

\bibitem[{Goicoechea \& Cernicharo(2001)}]{goicoechea2001far}
Goicoechea, J.~R., \& Cernicharo, J. 2001, The Astrophysical Journal Letters,
  554, L213

\bibitem[{Gordon {et~al.}(2016)Gordon, Rothman, Wilzewski, Kochanov, Hill, Tan,
  \& Wcislo}]{gordon2016hitran2016}
Gordon, I., Rothman, L.~S., Wilzewski, J.~S., {et~al.} 2016, in AAS/Division
  for Planetary Sciences Meeting Abstracts\# 48, Vol.~48

\bibitem[{Goyal {et~al.}(2017)Goyal, Mayne, Sing, Drummond, Tremblin, Amundsen,
  Evans, Carter, Spake, Baraffe, {et~al.}}]{goyal2017library}
Goyal, J.~M., Mayne, N., Sing, D.~K., {et~al.} 2017, Monthly Notices of the
  Royal Astronomical Society, 474, 5158

\bibitem[{Harps{\o}e {et~al.}(2013)Harps{\o}e, Hardis, Hinse, J{\o}rgensen,
  Mancini, Southworth, Alsubai, Bozza, Browne, Burgdorf,
  {et~al.}}]{harpsoe2013transiting}
Harps{\o}e, K. B.~W., Hardis, S., Hinse, T., {et~al.} 2013, Astronomy \&
  Astrophysics, 549, A10

\bibitem[{He {et~al.}(2018{\natexlab{a}})He, H{\"o}rst, Lewis, Yu, Moses,
  Kempton, McGuiggan, Morley, Valenti, \& Vuitton}]{he2018laboratory}
He, C., H{\"o}rst, S.~M., Lewis, N.~K., {et~al.} 2018{\natexlab{a}}, The
  Astrophysical Journal Letters, 856, L3

\bibitem[{He {et~al.}(2018{\natexlab{b}})He, H{\"o}rst, Lewis, Yu, Moses,
  Kempton, Marley, McGuiggan, Morley, Valenti, {et~al.}}]{he2018photochemical}
---. 2018{\natexlab{b}}, The Astronomical Journal, 156, 38

\bibitem[{He {et~al.}(2019)He, Hörst, Lewis, Moses, Kempton, Marley, Morley,
  Valenti, \& Vuitton}]{he2018gas}
He, C., Hörst, S.~M., Lewis, N.~K., {et~al.} 2019, ACS Earth and Space
  Chemistry, 3, 39

\bibitem[{Heays {et~al.}(2017)Heays, Bosman, \&
  Van~Dishoeck}]{heays2017photodissociation}
Heays, A., Bosman, AD, v., \& Van~Dishoeck, E. 2017, Astronomy \& Astrophysics,
  602, A105

\bibitem[{H{\'e}brard {et~al.}(2006)H{\'e}brard, Dobrijevic, B{\'e}nilan, \&
  Raulin}]{hebrard2006photochemical}
H{\'e}brard, E., Dobrijevic, M., B{\'e}nilan, Y., \& Raulin, F. 2006, Journal
  of Photochemistry and Photobiology C: Photochemistry Reviews, 7, 211

\bibitem[{H{\'e}brard {et~al.}(2009)H{\'e}brard, Dobrijevic, Pernot, Carrasco,
  Bergeat, Hickson, Canosa, Le~Picard, \& Sims}]{hebrard2009measurements}
H{\'e}brard, E., Dobrijevic, M., Pernot, P., {et~al.} 2009, The Journal of
  Physical Chemistry A, 113, 11227

\bibitem[{Helling \& Rimmer(2019)}]{helling2019lightning}
Helling, C., \& Rimmer, P.~B. 2019, arXiv preprint arXiv:1903.04565

\bibitem[{Hill {et~al.}(2013)Hill, Yurchenko, \&
  Tennyson}]{hill2013temperature}
Hill, C., Yurchenko, S.~N., \& Tennyson, J. 2013, Icarus, 226, 1673

\bibitem[{Hollenbach {et~al.}(2012)Hollenbach, Kaufman, Neufeld, Wolfire, \&
  Goicoechea}]{hollenbach2012chemistry}
Hollenbach, D., Kaufman, M., Neufeld, D., Wolfire, M., \& Goicoechea, J. 2012,
  The Astrophysical Journal, 754, 105

\bibitem[{H{\"o}rst(2017)}]{horst2017titan}
H{\"o}rst, S.~M. 2017, Journal of Geophysical Research: Planets, 122, 432

\bibitem[{H{\"o}rst {et~al.}(2018)H{\"o}rst, He, Lewis, Kempton, Marley,
  Morley, Moses, Valenti, \& Vuitton}]{horst2018haze}
H{\"o}rst, S.~M., He, C., Lewis, N.~K., {et~al.} 2018, Nature Astronomy, 2, 303

\bibitem[{Hu(2015)}]{hu2015photochemistry}
Hu, R. 2015, in Planetary Exploration and Science: Recent Results and Advances
  (Springer), 291--308

\bibitem[{Hu \& Seager(2014)}]{hu2014photochemistry}
Hu, R., \& Seager, S. 2014, The Astrophysical Journal, 784, 63

\bibitem[{Huebner \& Mukherjee(2015)}]{huebner2015photoionization}
Huebner, W., \& Mukherjee, J. 2015, Planetary and Space Science, 106, 11

\bibitem[{Indriolo \& McCall(2013)}]{indriolo2013cosmic}
Indriolo, N., \& McCall, B.~J. 2013, Chemical Society Reviews, 42, 7763

\bibitem[{Jontof-Hutter(2019)}]{jontof2019compositional}
Jontof-Hutter, D. 2019, Annual Review of Earth and Planetary Sciences, 47, 141

\bibitem[{Kempton {et~al.}(2011)Kempton, Zahnle, \&
  Fortney}]{kempton2011atmospheric}
Kempton, E. M.-R., Zahnle, K., \& Fortney, J.~J. 2011, The Astrophysical
  Journal, 745, 3

\bibitem[{Koskinen {et~al.}(2007)Koskinen, Aylward, Smith, \&
  Miller}]{koskinen2007thermospheric}
Koskinen, T., Aylward, A., Smith, C., \& Miller, S. 2007, The Astrophysical
  Journal, 661, 515

\bibitem[{Koskinen {et~al.}(2013)Koskinen, Harris, Yelle, \&
  Lavvas}]{koskinen2013escape}
Koskinen, T., Harris, M., Yelle, R., \& Lavvas, P. 2013, Icarus, 226, 1678

\bibitem[{Koskinen {et~al.}(2010)Koskinen, Yelle, Lavvas, \&
  Lewis}]{koskinen2010characterizing}
Koskinen, T., Yelle, R., Lavvas, P., \& Lewis, N. 2010, The Astrophysical
  Journal, 723, 116

\bibitem[{Kreidberg {et~al.}(2014)Kreidberg, Bean, D{\'e}sert, Benneke, Deming,
  Stevenson, Seager, Berta-Thompson, Seifahrt, \&
  Homeier}]{kreidberg2014clouds}
Kreidberg, L., Bean, J.~L., D{\'e}sert, J.-M., {et~al.} 2014, Nature, 505, 69

\bibitem[{Lammer {et~al.}(2019)Lammer, Spro{\ss}, Grenfell, Scherf, Fossati,
  Lendl, \& Cubillos}]{lammer2019role}
Lammer, H., Spro{\ss}, L., Grenfell, J.~L., {et~al.} 2019, arXiv preprint
  arXiv:1904.11716

\bibitem[{Lammer {et~al.}(2011)Lammer, Eybl, Kislyakova, Weingrill,
  Holmstr{\"o}m, Khodachenko, Kulikov, Reiners, Leitzinger, Odert,
  {et~al.}}]{lammer2011uv}
Lammer, H., Eybl, V., Kislyakova, K., {et~al.} 2011, Astrophysics and Space
  Science, 335, 39

\bibitem[{Larsson \& Orel(2008)}]{larsson2008dissociative}
Larsson, M., \& Orel, A.~E. 2008, Dissociative recombination of molecular ions
  (Cambridge University Press Cambridge)

\bibitem[{Laughlin {et~al.}(2004)Laughlin, Bodenheimer, \&
  Adams}]{laughlin2004core}
Laughlin, G., Bodenheimer, P., \& Adams, F.~C. 2004, The Astrophysical Journal
  Letters, 612, L73

\bibitem[{Lavie {et~al.}(2017)Lavie, Mendon{\c{c}}a, Mordasini, Malik,
  Bonnefoy, Demory, Oreshenko, Grimm, Ehrenreich, \& Heng}]{lavie2017helios}
Lavie, B., Mendon{\c{c}}a, J.~M., Mordasini, C., {et~al.} 2017, The
  Astronomical Journal, 154, 91

\bibitem[{Lavvas {et~al.}(2014)Lavvas, Koskinen, \& Yelle}]{lavvas2014electron}
Lavvas, P., Koskinen, T., \& Yelle, R.~V. 2014, The Astrophysical Journal, 796,
  15

\bibitem[{Lenz {et~al.}(2016)Lenz, Reiners, Seifahrt, \&
  K{\"a}ufl}]{lenz2016crires}
Lenz, L., Reiners, A., Seifahrt, A., \& K{\"a}ufl, H.-U. 2016, Astronomy \&
  Astrophysics, 589, A99

\bibitem[{Loyd {et~al.}(2016)Loyd, France, Youngblood, Schneider, Brown, Hu,
  Linsky, Froning, Redfield, Rugheimer, {et~al.}}]{loyd2016muscles}
Loyd, R.~P., France, K., Youngblood, A., {et~al.} 2016, The Astrophysical
  Journal, 824, 102

\bibitem[{Luger \& Barnes(2015)}]{luger2015extreme}
Luger, R., \& Barnes, R. 2015, Astrobiology, 15, 119

\bibitem[{Melnikov {et~al.}(2016)Melnikov, Yurchenko, Tennyson, \&
  Jensen}]{melnikov2016radiative}
Melnikov, V.~V., Yurchenko, S.~N., Tennyson, J., \& Jensen, P. 2016, Physical
  Chemistry Chemical Physics, 18, 26268

\bibitem[{Miguel {et~al.}(2014)Miguel, Kaltenegger, Linsky, \&
  Rugheimer}]{miguel2014effect}
Miguel, Y., Kaltenegger, L., Linsky, J.~L., \& Rugheimer, S. 2014, Monthly
  Notices of the Royal Astronomical Society, 446, 345

\bibitem[{Miller {et~al.}(2010)Miller, Stallard, Melin, \&
  Tennyson}]{miller2010h}
Miller, S., Stallard, T., Melin, H., \& Tennyson, J. 2010, Faraday discussions,
  147, 283

\bibitem[{Miller {et~al.}(2013)Miller, Stallard, Tennyson, \&
  Melin}]{miller2013cooling}
Miller, S., Stallard, T., Tennyson, J., \& Melin, H. 2013, The Journal of
  Physical Chemistry A, 117, 9770

\bibitem[{Miller {et~al.}(2000)Miller, Rego, Achilleos, Stallard, Prang{\'e},
  Dougherty, Joseph, Tennyson, Aylward, Meuller-Wodarg,
  {et~al.}}]{miller2000infrared}
Miller, S., Rego, D., Achilleos, N., {et~al.} 2000, Advances in Space Research,
  26, 1477

\bibitem[{Mizus {et~al.}(2017)Mizus, Alijah, Zobov, Lodi, Kyuberis, Yurchenko,
  Tennyson, \& Polyansky}]{mizus2017exomol}
Mizus, I.~I., Alijah, A., Zobov, N.~F., {et~al.} 2017, Monthly Notices of the
  Royal Astronomical Society, 468, 1717

\bibitem[{Moran {et~al.}(2020)Moran, Horst, Vuitton, He, Lewis, Flandinet,
  Moses, Orthous-Daunay, Sebree, \& Wolters}]{moran2020chemistry}
Moran, S., Horst, S., Vuitton, V., {et~al.} 2020, AAS, 52, 248

\bibitem[{Morley {et~al.}(2013)Morley, Fortney, Kempton, Marley, Vissher, \&
  Zahnle}]{morley2013quantitatively}
Morley, C.~V., Fortney, J.~J., Kempton, E. M.-R., {et~al.} 2013, The
  Astrophysical Journal, 775, 33

\bibitem[{Moses {et~al.}(2011)Moses, Visscher, Fortney, Showman, Lewis,
  Griffith, Klippenstein, Shabram, Friedson, Marley,
  {et~al.}}]{moses2011disequilibrium}
Moses, J.~I., Visscher, C., Fortney, J.~J., {et~al.} 2011, The Astrophysical
  Journal, 737, 15

\bibitem[{Moses {et~al.}(2013)Moses, Line, Visscher, Richardson, Nettelmann,
  Fortney, Barman, Stevenson, \& Madhusudhan}]{moses2013compositional}
Moses, J.~I., Line, M.~R., Visscher, C., {et~al.} 2013, The Astrophysical
  Journal, 777, 34

\bibitem[{Mugnai {et~al.}(2019)Mugnai, Edwards, Papageorgiou, Pascale, \&
  Sarkar}]{mugnaiarielrad}
Mugnai, L., Edwards, B., Papageorgiou, A., Pascale, E., \& Sarkar, S. 2019,
  EPSC-DPS Joint Meeting 2019, 13

\bibitem[{Mu{\~n}oz(2007)}]{munoz2007physical}
Mu{\~n}oz, A.~G. 2007, Planetary and Space Science, 55, 1426

\bibitem[{Neale {et~al.}(1996)Neale, Miller, \&
  Tennyson}]{neale1996spectroscopic}
Neale, L., Miller, S., \& Tennyson, J. 1996, The Astrophysical Journal, 464,
  516

\bibitem[{Owen \& Mohanty(2016)}]{owen2016habitability}
Owen, J.~E., \& Mohanty, S. 2016, Monthly Notices of the Royal Astronomical
  Society, 459, 4088

\bibitem[{Peng {et~al.}(2014)Peng, Carrasco, \& Pernot}]{peng2014modeling}
Peng, Z., Carrasco, N., \& Pernot, P. 2014, GeoResJ, 1, 33

\bibitem[{Pierrehumbert \& Gaidos(2011)}]{pierrehumbert2011hydrogen}
Pierrehumbert, R., \& Gaidos, E. 2011, The Astrophysical Journal Letters, 734,
  L13

\bibitem[{Plessis {et~al.}(2012)Plessis, Carrasco, Dobrijevic, \&
  Pernot}]{plessis2012production}
Plessis, S., Carrasco, N., Dobrijevic, M., \& Pernot, P. 2012, Icarus, 219, 254

\bibitem[{Polyansky {et~al.}(2018)Polyansky, Kyuberis, Zobov, Tennyson,
  Yurchenko, \& Lodi}]{polyansky2018exomol}
Polyansky, O.~L., Kyuberis, A.~A., Zobov, N.~F., {et~al.} 2018, Monthly Notices
  of the Royal Astronomical Society, 480, 2597

\bibitem[{Ribas {et~al.}(2016)Ribas, Bolmont, Selsis, Reiners, Leconte,
  Raymond, Engle, Guinan, Morin, Turbet, {et~al.}}]{ribas2016habitability}
Ribas, I., Bolmont, E., Selsis, F., {et~al.} 2016, Astronomy \& Astrophysics,
  596, A111

\bibitem[{Rimmer \& Helling(2016)}]{rimmer2016chemical}
Rimmer, P.~B., \& Helling, C. 2016, The Astrophysical Journal Supplement
  Series, 224, 9

\bibitem[{Rimmer \& Rugheimer(2019)}]{rimmer2019hydrogen}
Rimmer, P.~B., \& Rugheimer, S. 2019, Icarus, 329, 124

\bibitem[{Rimmer {et~al.}(2018)Rimmer, Xu, Thompson, Gillen, Sutherland, \&
  Queloz}]{rimmer2018origin}
Rimmer, P.~B., Xu, J., Thompson, S.~J., {et~al.} 2018, Science advances, 4,
  eaar3302

\bibitem[{Rothman {et~al.}(2010)Rothman, Gordon, Barber, Dothe, Gamache,
  Goldman, Perevalov, Tashkun, \& Tennyson}]{HITEMP}
Rothman, L.~S., Gordon, I.~E., Barber, R.~J., {et~al.} 2010, J. Quant.
  Spectrosc. Radiat. Transf., 111, 2139

\bibitem[{Scalo {et~al.}(2007)Scalo, Kaltenegger, Segura, Fridlund, Ribas,
  Kulikov, Grenfell, Rauer, Odert, Leitzinger, {et~al.}}]{scalo2007m}
Scalo, J., Kaltenegger, L., Segura, A., {et~al.} 2007, Astrobiology, 7, 85

\bibitem[{Schaefer {et~al.}(2012)Schaefer, Lodders, \&
  Fegley}]{schaefer2012vaporization}
Schaefer, L., Lodders, K., \& Fegley, B. 2012, The Astrophysical Journal, 755,
  41

\bibitem[{Shampine {et~al.}(2006)Shampine, Sommeijer, \&
  Verwer}]{shampine2006irkc}
Shampine, L.~F., Sommeijer, B.~P., \& Verwer, J.~G. 2006, Journal of
  computational and applied mathematics, 196, 485

\bibitem[{Shkolnik {et~al.}(2006)Shkolnik, Gaidos, \&
  Moskovitz}]{shkolnik2006no}
Shkolnik, E., Gaidos, E., \& Moskovitz, N. 2006, The Astronomical Journal, 132,
  1267

\bibitem[{Shkolnik \& Barman(2014)}]{shkolnik2014hazmat}
Shkolnik, E.~L., \& Barman, T.~S. 2014, The Astronomical Journal, 148, 64

\bibitem[{Stallard {et~al.}(2001)Stallard, Miller, Millward, \&
  Joseph}]{stallard2001dynamics}
Stallard, T., Miller, S., Millward, G., \& Joseph, R.~D. 2001, Icarus, 154, 475

\bibitem[{Tarter {et~al.}(2007)Tarter, Backus, Mancinelli, Aurnou, Backman,
  Basri, Boss, Clarke, Deming, Doyle, {et~al.}}]{tarter2007reappraisal}
Tarter, J.~C., Backus, P.~R., Mancinelli, R.~L., {et~al.} 2007, Astrobiology,
  7, 30

\bibitem[{Tennyson {et~al.}(2016)Tennyson, Yurchenko, Al-Refaie, Barton, Chubb,
  Coles, Diamantopoulou, Gorman, Hill, Lam, {et~al.}}]{tennyson2016exomol}
Tennyson, J., Yurchenko, S.~N., Al-Refaie, A.~F., {et~al.} 2016, Journal of
  Molecular Spectroscopy, 327, 73

\bibitem[{Tian(2009)}]{tian2009thermal}
Tian, F. 2009, The Astrophysical Journal, 703, 905

\bibitem[{Tigrine {et~al.}(2016)Tigrine, Carrasco, Vettier, \&
  Cernogora}]{tigrine2016microwave}
Tigrine, S., Carrasco, N., Vettier, L., \& Cernogora, G. 2016, Journal of
  Physics D: Applied Physics, 49, 395202

\bibitem[{Tinetti {et~al.}(2016)Tinetti, Drossart, Eccleston, Hartogh, Heske,
  Leconte, Micela, Ollivier, Pilbratt, Puig, {et~al.}}]{tinetti2016science}
Tinetti, G., Drossart, P., Eccleston, P., {et~al.} 2016, in Space Telescopes
  and Instrumentation 2016: Optical, Infrared, and Millimeter Wave, Vol. 9904,
  International Society for Optics and Photonics, 99041X

\bibitem[{Tinetti {et~al.}(2018)Tinetti, Drossart, Eccleston, Hartogh, Heske,
  Leconte, Micela, Ollivier, Pilbratt, Puig, {et~al.}}]{tinetti2018chemical}
Tinetti, G., Drossart, P., Eccleston, P., {et~al.} 2018, Experimental
  Astronomy, 46, 135

\bibitem[{Trafton {et~al.}(1993)Trafton, Geballe, Miller, Tennyson, \&
  Ballester}]{trafton1993detection}
Trafton, L.~M., Geballe, T., Miller, S., Tennyson, J., \& Ballester, G. 1993,
  The Astrophysical Journal, 405, 761

\bibitem[{Van~Dishoeck {et~al.}(2013)Van~Dishoeck, Herbst, \&
  Neufeld}]{van2013interstellar}
Van~Dishoeck, E.~F., Herbst, E., \& Neufeld, D.~A. 2013, Chemical Reviews, 113,
  9043

\bibitem[{Venot {et~al.}(2014)Venot, Ag{\'u}ndez, Selsis, Tessenyi, \&
  Iro}]{venot2014atmospheric}
Venot, O., Ag{\'u}ndez, M., Selsis, F., Tessenyi, M., \& Iro, N. 2014,
  Astronomy \& Astrophysics, 562, A51

\bibitem[{Verwer \& Sommeijer(2004)}]{verwer2004implicit}
Verwer, J.~G., \& Sommeijer, B.~P. 2004, SIAM Journal on Scientific Computing,
  25, 1824

\bibitem[{Vuitton {et~al.}(2019{\natexlab{a}})Vuitton, Flandinet,
  Orthous-Daunay, Cedric, Ayoub, Horst, He, \& Moran}]{vuitton2019laboratory}
Vuitton, V., Flandinet, L., Orthous-Daunay, F.-R., {et~al.} 2019{\natexlab{a}},
  in 2019 Astrobiology Science Conference, AGU

\bibitem[{Vuitton {et~al.}(2019{\natexlab{b}})Vuitton, Yelle, Klippenstein,
  H{\"o}rst, \& Lavvas}]{vuitton2019simulating}
Vuitton, V., Yelle, R., Klippenstein, S., H{\"o}rst, S., \& Lavvas, P.
  2019{\natexlab{b}}, Icarus, 324, 120

\bibitem[{Wakelam {et~al.}(2012)Wakelam, Herbst, Loison, Smith, Chandrasekaran,
  Pavone, Adams, Bacchus-Montabonel, Bergeat, B{\'e}roff,
  {et~al.}}]{wakelam2012kinetic}
Wakelam, V., Herbst, E., Loison, J.-C., {et~al.} 2012, The Astrophysical
  Journal Supplement Series, 199, 21

\bibitem[{Waldmann {et~al.}(2015{\natexlab{a}})Waldmann, Rocchetto, Tinetti,
  Barton, Yurchenko, \& Tennyson}]{waldmann2015rex}
Waldmann, I.~P., Rocchetto, M., Tinetti, G., {et~al.} 2015{\natexlab{a}}, The
  Astrophysical Journal, 813, 13

\bibitem[{Waldmann {et~al.}(2015{\natexlab{b}})Waldmann, Tinetti, Rocchetto,
  Barton, Yurchenko, \& Tennyson}]{waldmann2015tau}
Waldmann, I.~P., Tinetti, G., Rocchetto, M., {et~al.} 2015{\natexlab{b}}, The
  Astrophysical Journal, 802, 107

\bibitem[{Wolters {et~al.}(2019)Wolters, Vuitton, Flandinet, Moran, He, Ayoub,
  Orthous-Daunay, \& H{\"o}rst}]{wolters2019orbitrap}
Wolters, C., Vuitton, V., Flandinet, L., {et~al.} 2019, EPSC, 2019, EPSC

\bibitem[{Wordsworth \& Pierrehumbert(2014)}]{wordsworth2014abiotic}
Wordsworth, R., \& Pierrehumbert, R. 2014, The Astrophysical Journal Letters,
  785, L20

\bibitem[{Yelle(2004)}]{yelle2004aeronomy}
Yelle, R.~V. 2004, Icarus, 170, 167

\bibitem[{Youngblood {et~al.}(2016)Youngblood, France, Loyd, Linsky, Redfield,
  Schneider, Wood, Brown, Froning, Miguel, {et~al.}}]{youngblood2016muscles}
Youngblood, A., France, K., Loyd, R.~P., {et~al.} 2016, The Astrophysical
  Journal, 824, 101

\bibitem[{Yurchenko {et~al.}(2014)Yurchenko, Tennyson, Bailey, Hollis, \&
  Tinetti}]{yurchenko2014spectrum}
Yurchenko, S.~N., Tennyson, J., Bailey, J., Hollis, M.~D., \& Tinetti, G. 2014,
  Proceedings of the National Academy of Sciences, 111, 9379

\end{thebibliography}
\bibliographystyle{aasjournal}



\end{document}